\documentclass[10pt]{article}

\usepackage{graphicx}
\usepackage[colorinlistoftodos]{todonotes}
\usepackage{amsfonts}
\usepackage{amsmath}
\usepackage{subcaption}

\newcommand{\kernel}[1]{\ensuremath{\left.\left<q\right|#1\left|q'\right>\right.}}
\newcommand{\kernelneg}[1]{\ensuremath{\left.\left<-q\right|#1\left|-q'\right>\right.}}
\newcommand{\kernelpos}[1]{\ensuremath{\left.\left<q\right|#1\left|q'\right>\right.}}
\newcommand{\kernelc}[1]{\ensuremath{\left.\left<q'\right|#1\left|q\right>\right.}}
\newcommand{\transker}[1]{\ensuremath{\left.\left<q+\frac{x}{2}\right|#1\left|q-\frac{x}{2}\right>\right.}}
\newcommand{\opr}[1]{\ensuremath{\widehat{\mathrm{#1}}}}
\newcommand{\kernelqq}[3]{\ensuremath{\left<#1\left|#2\right|#3\right>}}
\newcommand{\ket}[1]{\ensuremath{\left|\left. #1\right>\right.}}
\newcommand{\bra}[1]{\ensuremath{\left<\left. #1\right|\right.}}
\newcommand{\braket}[2]{\ensuremath{\left<\left.#1\left| #2\right>\right.\right.}}

\newcommand{\dd}[1]{\ensuremath{\mathrm{d}#1}}
\newcommand{\sgn}[1]{\ensuremath{\mathrm{sgn}\left(#1\right)}}
\newcommand{\kernell}[1]{\ensuremath{\left.\left<q'\right|#1\left|q\right>\right.}}

\begin{document}

\title{Quantizations of the classical time of arrival and their dynamics}
\author{Eric A. Galapon\thanks{eric.galapon@up.edu.ph} and John Jaykel P. Magadan\\Theoretical Physics Group, National Institute of Physics\\University of the Philippines, 1101 Philippines}

\maketitle
\begin{abstract}
The classical time of arrival in the interacting case is quantized by way of quantizing its expansion about the free time of arrival. The quantization is formulated in coordinate representation which represents ordering rules in terms of two variable polynomial functions. This leads to representations of the quantized time of arrival operators as integral operators whose kernels are determined by the chosen ordering rule. The formulation lends itself to generalization which allows construction of time of arrival operators that cannot be obtained by direct quantization using particular ordering rules. Wey, symmetric and Born-Jordan quantizations are specifically studied. The dynamics of the eigenfunctions of the different time of arrival operators are investigated. The eigenfunctions exhibit unitary arrival at the intended arrival point at their respective eigenvalues. 
\end{abstract}
\section{Introduction}
Observables of a quantum system are represented by hermitian operators in the underlying state space or Hilbert space of the system. A quantum observable is a quantum image of a classical observable if the corresponding operator has a classical limit which is a real valued function of the classical state. For a massive and structureless particle, its classical state is a point in the position-momentum phase space and its classical observables are real valued functions of its position and momentum. A quantum image of any of the particle's classical observable is typically obtained by quantization \cite{degoson,degosonbook,degosonPR,ali,bender,domingoJMP2015,tosiek}.
Quantization is implemented by promoting the classical position and momentum observables into non-commuting operators that satisfy the canonical commutation relation. However, the non-commutativity of the position and momentum operators raises ordering ambiguity in quantization, leading to potentially infinitely many possible quantizations of the classical observable corresponding to the infinitely many possible orderings of the non-commuting operators. Of the many possible quantizations, Weyl, simple symmetric ordering and Born-Jordan quantizations are the most employed and studied quantizations \cite{degoson,degosonbook,degosonPR,tosiek}. Quantization already make it clear that the quantum image of a given classical observable comprises a large set. But quantization does not exhaust all possible quantum images of a given classical observable. A quantum image can be constructed without quantization and constructed only from quantum first principles, for example, Mackey's construction of the position and momentum operators without quantization and one of us' construction of a quantum time of arrival operator without quantization \cite{mac1,mac2}. The fundamental question is whether there is one distinguished quantum image over the others or all quantum images are on equal footing for a given classical observable \cite{degosonbook}. 

In this paper we will consider quantizations of the classical time of arrival of a structureless particle in one dimension under the influence of an interaction potential $V(q)$ and compare these quantizations \cite{leon,villanuevaPRA2010,galaponIJMP2006,villanuevaJPA2008,galaponJMP2004,sombilloJMP2012,peresbook,baute,mug21,galaponPRSA2009,galaponOS2001}. For such a particle with initial position and momentum $q$ and $p$, respectively, at $t=0$, the time of arrival at $x$ in the configuration space is given by 
\begin{equation}\label{prob}
T_x(q,p)=-\mbox{sgn}(p) \sqrt{\frac{\mu}{2}}\int_x^q \frac{dq'}{\sqrt{H(q,p)-V(q')}},
\end{equation}
where $H(q,p)$ is the Hamiltonian or the initial total energy of the particle,
\begin{equation}
H(q,p)=\frac{p^2}{2\mu} +V(q),
\end{equation}
in which $\mu$ is the mass of the particle and $\mbox{sgn}(p)$ is the sign of the initial momentum $p$. It is expression $T_x(q,p)$ given by equation \eqref{prob} that we seek to quantize.

However, the quantization of time has been a contentious issue and the problem of non-parametric or dynamical nature of time in quantum mechanics remains lacking in consensus \cite{pauli,hauge,landauer,timev1,timev2,mug5,galaponSpringer,galaponWorld}. Arguments against the existence of meaningful time operator has been the non-existence of self-adjoint covariant time operator for discrete or semi-bounded Hamiltonians \cite{pauli}. This has been the problem for the free quantum time of arrival problem \cite{allcock,allcock2,allcock3,kijowski,muga2,muga1,leavens,toa4,toa1,galaponPRA2005a,caballar2,caballarPRA2010,galaponPRA2005b,sombilloAP2016,galaponPRA2009}. But this issue has been adequately addressed by the clarification of the long standing non-go theorem of Pauli for the existence of self-adjoint operators \cite{galaponPRSA2002,galaponPRSA2002b} and the introduction of non-projective observables in quantum mechanics \cite{mug5,busch,busch2,srinivas,amann}. But the same cannot be said in the interacting case. By inspection of the classical expression $T_x(q,p)$, we find that it can be real, signifying that the arrival point $x$ is accessible to the particle given its initial state $(q,p)$ in the phase space; it can be complex or infinity, signifying that the arrival point is not accessible to the particle; moreover, it can be multiple valued when it is real. That is $T_x(q,p)$ is not defined in the entire phase space contrary to the free particle time of arrival $t=-\mu (q-x)/p$ which is real and single valued everywhere. It is for these reasons that it had been deemed not meaningful to quantize the classical time of arrival in the interacting case because the expression is generally not real valued in the entire phase space and at the same time it may be multiple valued \cite{leon,peresbook}.

But these objections can be adequately addressed on physical grounds. First, we know that the entire phase space is accessible to the quantum particle, thanks to the quantum tunneling effect. This implies that a quantum particle may be localized in a forbidden region in the phase space but it can tunnel to the arrival point even though it does not have sufficient energy. This means that in this case it is meaningful to ask of the time of arrival. Second, when the classical time of arrival is multiple valued, the same cannot be said to the corresponding quantum observable, assuming it exists. The reason follows from the difference in the nature of measurements in classical and quantum mechanics. In classical mechanics, we can measure in principle without disturbing the system. This allows us to observe multiple arrivals and see the periodicity inherent in the classical expression. However, that cannot be said in the quantum case. In quantum mechanics, measurement irreversibly changes the state of the system. This means that the state of the system after the registration of the first arrival is no longer causally related to the initial state of the system or even just before the measurement was made. This means that if another arrival is registered after the first, the second cannot be interpreted as the second arrival starting from the initial state of the system. For this reason only the first arrival may only be meaningful to ask if we insist that arrival is interpreted as a detection at the arrival point. Hence the known laws of quantum mechanics does not at all forbid the quantization of the classical time of arrival. The only question now is how to proceed with the quantization. 

Earlier one of us considered the problem of constructing a time of arrival operator without quantization \cite{galaponJMP2004}. The operator was constructed only on pure quantum mechanical considerations, with the classical observable serving as a boundary condition, contrary to quantization which served as the starting point. We referred to the construction of a quantum observable without quantization as supraquantization, and the operator constructed as a supraquantized operator. There we obtained the supraquantized time of arrival (STOA) operator, $\opr{T}$, under four conditions: (i) it is hermitian, $\opr{T}^{\dagger}=\opr{T}$; (ii) it satisfies the time reversal symmetry $\opr{\Theta}\opr{T}\opr{\Theta}^{\dagger}=-\opr{T}$, where $\opr{\Theta}$ is the time reversal operator, (iii) it is conjugate to the Hamiltonian $\opr{H}$, $([\opr{H},\opr{T}]\varphi)(q)=i\hbar\varphi(q)$; and (iii) it reduces to the classical time of arrival in the classical limit as $\hbar\rightarrow 0$.  The STOA-operator appears as an integral operator in configuration space,
\begin{equation}\label{eq:eigvalue}
(\opr{T}\varphi)(q)=\int_{-\infty}^{\infty}\kernel{\opr{T}} \varphi(q')\,\mathrm{d}q'
\end{equation}
where the kernel is given by
\begin{eqnarray}\label{eq:T}
\kernel{\opr{T}} = \frac{\mu}{i\hbar}T(q,q')\mbox{sgn}(q-q')
\end{eqnarray}
and $T(q,q')$, which we refer to as the kernel factor, depends on the Hamiltonian, and is required to be real-valued, symmetric and analytic for the STOA-operator, $\opr{T}$, to satisfy hermicity and time reversal symmetry requirements. The hermicity condition on $\opr{T}$ translates to $\kernel{\opr{T}}=\kernell{\opr{T}}^*$; while the time reversal symmetry requirement, to$\kernel{T}^*=-\kernel{\opr{T}}$. Thus when $T(q,q')$ is real valued both requirements are satisfied.

The kernel factor is the solution to the second order partial differential equation
\begin{eqnarray}\label{eq:timeKernelDifferentialEqn}
- \frac{\hbar^{2}}{2\mu} \frac{\partial^{2} T(q,q')}{\partial q^{2}} +\frac{\hbar^{2}}{2\mu}\frac{\partial^{2} T(q,q')}{\partial q'^{2}} +\Bigr (V(q)-V(q') \Bigr)T(q,q')=0
\end{eqnarray}
subject to the boundary conditions
\begin{eqnarray}\label{eq:boundaryConditions}
T(q,q)=\frac{q}{2}, \;\;\;\;
T(q,-q)=0.
\end{eqnarray}
The partial differential equation \eqref{eq:timeKernelDifferentialEqn} is a direct consequence of the conjugacy requirement; on the other hand, the boundary conditions along the diagonals given by \eqref{eq:boundaryConditions} ensure that the time of arrival operator has the required classical limit. We referred to equation (\ref{eq:timeKernelDifferentialEqn}) as the time kernel equation (TKE) \cite{galaponJMP2004}. It can be established that the TKE, under the stated conditions, admits a unique solution for continuous interaction potentials \cite{sombilloJMP2012,domingoMS}. 

The SQTOA-operator $\opr{T}$ is not the only operator that is canonically conjugate with the Hamiltonian $\opr{H}$. Given $\opr{T}$, the operator $\opr{T}'=\opr{T}+f[
\opr{H}]$, for some sufficiently smooth function $f$, is conjugate with $\opr{H}$. However, while $f$ can be chosen such that $\opr{T}'$ is hermitian, in general for any $f$, $\opr{T}'$ will not satisfy the time reversal symmetry requirement because the Hamiltonian is invariant under time reversal. It has been establish in \cite{caballarPRA2010} that the absence of the required time reversal symmetry has a profound negative effect on the required dynamics of a time of arrival operator, which is unitary arrival of the TOA-operator eigenfunctions at the intended arrival point at their respective eigenvalues \cite{galaponPRA2005a,caballar2}. The operator $\opr{T}'$, while satisfying conjugacy with the Hamiltonian and may be constructed to satisfy the required classical limit, it cannot be considered as a time of arrival operator because of the non-arrival of their eigenfunctions at their eigenvalues. The SQTOA-operator is distinguished from the other solutions to the time-energy commutation relation due to its uniqueness arising from the condition \eqref{eq:boundaryConditions} which is the translation of the quantum-classical correspondence principle applied to the time of arrival observable, in addition to the fact that it satisfies the hermicity, time-reversal symmetry and conjugacy with the Hamiltonian.\footnote{An explicit operator formulation of the problem of constructing conjugate time of arrival operators is developed in \cite{villanuevaJPA2008} which is based on the methods of \cite{benderPRD1989a,benderPRD1989b}. Both \cite{villanuevaJPA2008} and \cite{benderPRD1989a,benderPRD1989b} solve the problem of finding an operator conjugate to a given Hamiltonian, but they differ in that \cite{villanuevaJPA2008} specifically solves the operator that reduces to the classical time of arrival in the classical limit while \cite{benderPRD1989a,benderPRD1989b} solves for the operator that minimally satisfies the canonical commutation relation.} 

Now an important property of the solution to the time kernel equation is that, for continuous potentials, it admits the expansion
$T(q,q')=T_0(q,q') + T_1(q,q')+ T_2(q,q') + \dots $,
so that the kernel of the time of arrival operator likewise assumes the expansion
$\kernel{\opr{T}}=\kernel{\opr{T}_0}+\kernel{\opr{T}_1}+ \kernel{\opr{T}_2}+\dots $,
where 
$\kernel{\opr{T}_k}=\frac{\mu}{i\hbar} T_k(q,q')\mbox{sgn}(q-q')$; equivalently, the STOA-opertor assumes the expansion $\opr{T}=\opr{T}_0 + \opr{T}_1 +\opr{T}_2 + \dots$. We find that the Weyl-Wigner transform of the kernel $\kernel{\opr{T}}$ is given by
\begin{eqnarray}
\mathcal{T}_{\hbar}(q,p)&=&\int_{-\infty}^{\infty}\transker{\opr{T}} e^{-i x p/\hbar} \, \mathrm{d}x \nonumber \\
&=&\sum_{k=0}^{\infty} \frac{\mu}{i\hbar} \int_{-\infty}^{\infty} T_k\left(q+\frac{x}{2},q-\frac{x}{2}\right) \mbox{sgn}(x) e^{- i x p/\hbar} \, \mathrm{d}x \nonumber \\
&=&\tau_0(q,p)+\hbar^2 \tau_1(q,p) + \hbar^4 \tau_2(q,p)+\dots ,\label{bebe}
\end{eqnarray}
where each $\tau_k(q,p)$ is independent of $\hbar$. 

The leading term $\tau_0(q,p)$ turns out to be just the expansion of equation \eqref{prob} about the free time of arrival and is given by 
\begin{equation}\label{ltoa}
\tau(x;q,p)=-\sum_{k=0}^{\infty}(-1)^k \frac{(2k-1)!!}{k!} \frac{\mu^{k+1}}{p^{2k+1}}\int_x^q (V(q)-V(q'))^k \,dq' .
\end{equation}
We referred to this as the local time of arrival (LTOA) \cite{galaponJMP2004}. Clearly from  equation \eqref{bebe} the STOA-operator has the LTOA as its classical limit, and the full classical time arrival \eqref{prob} by extension. Most important is the realization that the leading term in the expansion of the STOA-operator, $\opr{T}_0$, is the Weyl-quantization of the LTOA. That is, for arrival at the origin, the leading kernel in the expansion of the full kernel, $\kernel{\opr{T}}$, is given by
\begin{equation}
\kernel{\opr{T}_0}=\int_{-\infty}^{\infty}\tau\!\left(x=0; \frac{q+q'}{2},p\right) e^{i (q-q') p/\hbar} \, \frac{\mathrm{d}p}{2\pi\hbar} ,
\end{equation}
in accordance with Weyl quantization. For linear systems, that is systems with interaction potentials $V(q)=a + b q + c q^2$ where $a$, $b$ and $c$ are constants, the $T_k(q,q')$'s identically vanish, $T_k(q,q')=0$, for all $k=1,2,3, \dots$ so that the supraquantized time of arrival operator is given by $\kernel{\opr{T}}=\kernel{\opr{T}_0}$. That is $\opr{T}_0$ is just the Weyl quantization of the local time of arrival at the origin. On the other hand, for non-linear systems, the $T_k(q,q')$'s do not vanish for all $k$, with the leading term, again, being just the Weyl quantization of the LTOA; while the leading term can be obtained by Weyl quantization, the rest of the terms in the expansion of the solution must be solved explicitly from the time kernel equation.

Now the LTOA is single and real valued within its region of convergence in the phase space, which is in the neighborhood of the (classical) free time of arrival. Since the leading term of the STOA-operator is the Wey-quantized LTOA, supraquantization suggests that if we wish to quantize the classical time of arrival it is the LTOA that we should quantize. In this paper, we will do just that---quantize the local time of arrival. In contrast to supraquantization, quantization does not provide a unique quantum image to the local time of arrival. In fact there are infinitely many quantizations corresponding to the uncountable possible ways of ordering the non-commuting position and momentum operators that appear in the expansion of the LTOA. Moreover, the quantized time of arrival operators will not necessarily be canonically conjugate with the system Hamiltonian. Quantization of the LTOA was already considered earlier but the quantization was only using Weyl quantization \cite{galaponIJMP2006,villanuevaJPA2008}. Here we will obtain the most general quantization of the LTOA. We will then numerically investigate their dynamics, in particular, the dynamics of the eigenfunctions of the quantized time of arrival operators in relation with their respective eigenvalues. The result will give insight on the role of the conjugacy with the observed earlier dynamics of the quantized LTOA using Weyl quantization. 

The paper is organized as follows. In Section-\ref{observables}, we discuss the position representation of quantum observables and their classical limit within the rigged Hilbert space formulation of quantum mechanics. In Section-\ref{quantization} we discuss quantization in position representation and the representation of operators as formal integral operators. There we will discuss the most general quantizations of the monomial $q^n p^{-m}$ for positive integers $n$ and $m$. In Section-\ref{qltoa} we perform the quantization of the local time of arrival operator and obtain the general form of the kernel for the quantized local time of arrival operator. Explicit forms of the kernels for Weyl, Born-Jordan and simple symmetric ordering are obtained. In Section-\ref{harmonic} we obtain the quantized LTOA-operator for the harmonic oscillator for Weyl, Born-Jordan and simple symmetric ordering. In Section-\ref{generalquantumimage} we outline the construction of the most general quantum image of the local time of arrival. In Section-\ref{parity} we derive under what condition the eigenfunctions of the quantized LTOA-operators have definite parities. In Section-\ref{dynamics} we study numerically the time development of the eigenfunctions of the quantized LTOA-operators to compare the dynamics of the different quantizations. In Section-\ref{conclusion} we conclude.

\section{Quantum observables in position representation and their classical limit}\label{observables}
We now consider a structureless classical particle in one-dimension. We will describe such a particle quantum mechanically within the rigged Hilbert space formulation of quantum mechanics \cite{raf1,raf2,raf3,bohm}. We choose our rigged Hilbert space to be $\Phi\subset L^2(\mathbb{R})\subset \Phi^{\times}$, where $\Phi$ is the fundamental space of infinitely continuously differentiable complex valued functions with compact supports. The standard Hilbert space formulation of quantum mechanics can be obtained from the rigged Hilbert space by taking closures in $\Phi$ with respect to the metric of $L^2({\mathbb{R}})$. The advantage in using the rigged Hilbert space is that it allows us represent observables as distributions which allows manipulations of observables as scalar objects. 

A quantum observable in RHS is a mapping $\opr{A}:\Phi\mapsto \Phi^{\times}$, and, in coordinate representation, it is given by the formal integral operator
\begin{equation}\label{operatorinposrep}
(\opr{A}\varphi)\!(q)=\int_{-\infty}^{\infty}\kernel{\opr{A}}\varphi(q')\,\mathrm{d}q',
\end{equation}
where the kernel satisfies the hermicity condition $\kernel{\opr{A}}=\kernelc{\opr{A}}^*$. Strictly speaking, equation \eqref{operatorinposrep} should be interpreted in the distributional sense with the kernel as a distribution and the integral as a functional on $\Phi$. For example, the position and momentum operators are given by
\begin{equation}
(\opr{q}\varphi)(q)  = q \varphi(q) = \int_{\infty}^{\infty} q' \delta(q-q')\, \varphi(q')\, \mathrm{d}q' 
\end{equation}
\begin{equation}
(\opr{p}\varphi)(q) = -i\hbar \varphi'(q) =\int_{-\infty}^{\infty} i \hbar \delta'(q-q')\, \varphi(q')\, \mathrm{d} q'
\end{equation}
In these examples, the kernels are distributions and the integral is to be understood as a functional over the space $\Phi$. In our present case, we will find that the operator is a legitimate integral operator with regular kernels.

Given a classical observable $A=A(q,p)$, its quantum image is typically constructed by quantization. For example, the Weyl quantization of $A$ is specfied by its kernel which is given by
\begin{equation}
\kernel{\opr{A}_W}=\int_{-\infty}^{\infty}A\left(\frac{q+q'}{2},p\right) e^{i (q-q') p/\hbar} \, \frac{\mathrm{d}p}{2\pi\hbar} .
\end{equation}
For the position operator, Weyl quantization yields the kernel
\begin{eqnarray}
\kernel{\opr{q}_W} = \int_{-\infty}^{\infty} \frac{(q+q')}{2} e^{i (q-q')p/\hbar}\frac{\mathrm{d}p}{2\pi\hbar} = q \delta(q-q')
\end{eqnarray}
obtained upon using the representation $2\pi \delta(x) = \int_{-\infty}^{\infty} e^{x t} \mathrm{d}t$ and the property $f(x)\delta(x-x_0)=f(x_0) \delta(x-x_0)$. This coincides with the above kernel. For the momentum operator,
\begin{eqnarray}
\kernel{\opr{p}_W} &=& \int_{-\infty}^{\infty} p e^{i (q-q') p/\hbar} \frac{\mathrm{d}p}{2\pi\hbar} \nonumber \\
&=& \frac{\hbar}{i} \frac{\mathrm{d}}{\mathrm{d} q} \int_{-\infty}^{\infty} e^{i (q-q')p/\hbar} \frac{\mathrm{d}p}{2\pi \hbar} \nonumber \\
&=& i\hbar \frac{\mathrm{d}}{\mathrm{d}q'}\delta(q-q')
\end{eqnarray}
which reproduces the above kernel for the momentum operator. The classical observable is recovered by means of the inverse Fourier transform,
\begin{equation}
A(q,p)=\int_{-\infty}^{\infty}\transker{\opr{A}_W} e^{-i x p/\hbar} \, \mathrm{d}x .
\end{equation}
Using the known properties of the Dirac delta function, the classical position and momentum observables are recovered from this expression.

In general, the quantum image of a classical observable $A=A(q,p)$ is a set of hermitian distributions in $\Phi^*$, $\{\kernel{\opr{A}}\}$. A large portion of this set arises from quantizations of $A(q,p)$ and the rest those that cannot be obtained by direct quantization reduces to $A(q,p)$ in the classical limit. Explicitly the classical limit is obtained by
\begin{equation}\label{climit}
A(q,p)=\lim_{\hbar\rightarrow 0}\int_{-\infty}^{\infty}\transker{\opr{A}} e^{-i x p/\hbar} \,  \dd{x} ,
\end{equation}
where the integral should be understood in the distributional sense, provided the limit exists \cite{weyl,deg}. Observe that classical limit of the Weyl quantization does not involve the limit of vanishing $\hbar$. This implies that the leading term in the classical limit \ref{climit} (without the limit being performed) is just the classical limit of the Weyl transformed operator itself. This implies further that, if expansion is possible on other quantizations, then the leading term is the Weyl quantized operator. This has already been observed in the supraquantization of the classical time of arrival done much earlier \cite{galaponJMP2004}. 

\section{Quantization of  the monomials $q^n p^{-m}$}\label{quantization}
For interaction potentials that are analytic at the origin, it will be shown below that the local time of arrival is an expansion in terms of the monomials $q^n p^{-m}$ for $n=0,1, 2, \dots$ and $m=1, 2, 3, \dots$. Since quantization is linear, the problem of quantizing the LTOA reduces to the problem of quantizing the $q^np^{-m}$'s. Here we consider the most general quantization of the monomials via analytic ordering rules. We define analytic ordering rules by way of an example: Consider the quantization of $q p^{-1}$. This scalar expression can be quantized in the form $\opr{p}^{-1/2}\opr{q}\opr{p}^{-1/2}$ \cite{toa4}; since the square-root is non-analytic, we refer to this as a non-analytic quantization of the given scalar quantity. But we can also have the quantization $(\opr{q}\opr{p}^{-1}+\opr{p}^{-1}\opr{q})/2$; since all functions involved, which are integer powers of $\opr{q}$ and $\opr{p}^{-1}$, are analytic, we refer to this as an analytic quantization of $q p^{-1}$.

\subsection{Operator ordering rules}
Imposing analytic quantization of $q^n p^{-m}$ restricts us to operator ordering rules that involve only integer powers of $\opr{q}$ and $\opr{p}^{-1}$. We generalize Bender and Dunne's quantization prescription in \cite{bender} to accomodate this case. In particular, we have the quantization 
\begin{eqnarray}\label{order}
	Q(q^{n}p^{-m}) \to \opr{M}^{(Q)}_{n,m}= \frac{\sum_{j=0}^{n} a^{(n)}_{j}\opr{q}^{j}\opr{p}^{-m}\hat{q}^{n-j}}{\sum_{j=0}^{n}a^{n}_{j}}
\end{eqnarray}
where the $a_j^{(n)}$'s satisfy the condition $a_k^{(n)}=\bar{a}^{(n)}_{n-k}$. Evidently the right hand side reduces to the classical expression in the limit of commuting position and momentum operators. We require that the $a_j^{(n)}$'s are free from any parameter and, specifically, independent of $\hbar$. 

Three of the well-known quantization formulas are symmetric ordering, Born-Jordan Ordering and Weyl Ordering. For symmetric ordering, the coefficients are given by $a^{(n)}_0 = a^{(n)}_n = 1$ and $a_k^{(n)} = 0$ for $0 < k <n$. For Born-Jordan Ordering, $a_k^{(n)}$ for all $ 0 \leq k \leq n$. For Weyl ordering, $a_k^{(n)}={n \choose k}$. Explicitly the these quantizations yield, respectively, the following operators,
\begin{equation}
\opr{M}_{n,m}^{(W)}=\opr{T}_{n,-m}=\frac{1}{2^n}\sum_{j=0}^{n} {n \choose j} \opr{q}^j \opr{p}^{-m}\opr{q}^{n-j},
\end{equation}
\begin{equation}
\opr{M}_{n,m}^{(S)}=\opr{S}_{n,-m}=\frac{1}{2}\left(\opr{q}^n\opr{p}^{-m}+\opr{p}^{-m} \opr{q}^n\right),
\end{equation}
\begin{equation}
\opr{M}_{n,m}^{(BJ)}=\opr{B}_{n,-m}=\frac{1}{n+1}\sum_{j=0}^n \opr{q}^j \opr{p}^{-m}\opr{q}^{n-j},
\end{equation}
where the symbols at the middle equality denote the standard notations for them. The set of operator quantizations $\{\opr{T}_{r,s}\}$, $\{\opr{S}_{r,s}\}$, $\{\opr{B}_{r,s}\}$ for integers $r$ and $s$ form a basis expansions for operators \cite{domingoJMP2015,benderPRD1989a,benderPRD1989b,bunaoJMP2014}.  

\subsection{The ordering rules in coordinate representation}

In coordinate representation, the quantized monomial is an integral operator whose kernel is given by
\begin{equation}\label{wewe}
	\kernel{\opr{M}_{n,m}^{(Q)}} = \frac{1}{\sum_{j=0}^{n}a^{(n)}_{j}} \sum_{j=0}^{n} a^{(n)}_{j} q^{j} q'^{n-j}\kernel{\opr{p}^{-m}}
\end{equation}
The kernel of $\opr{p}^{-m}$ is obtained by inserting the resolution of the identity $\opr{1}=\int_{-\infty}^{\infty} \ket{p}\!\!\bra{p}\dd{p}$, where $\ket{p}$ is an eigenket of $\opr{p}$, i.e., $\opr{p}\ket{p}=p\ket{p}$. Then
\begin{eqnarray}
\kernel{\opr{p}^{-m}} &=& \int_{-\infty}^{\infty} \frac{\braket{q}{p}\braket{p}{q'}}{p^m}\, \mathrm{d}p\nonumber \\
&=& \frac{1}{2\pi \hbar} \int_{-\infty}^{\infty} \frac{e^{i (q-q')p/\hbar}}{p^m} \mathrm{d}p\label{integral}
\end{eqnarray}
The second line follows from the fact that $\braket{q}{p}$ is just the free plane wave in position representation, $\braket{q}{p}=e^{i q p/\hbar}/\sqrt{2\pi\hbar}$.

To evaluate the integral $\ref{integral}$, we interpret it as a distributional integral, in particular a distributional Fourier transform. We use the known Fourier transform \cite[p.360,\#18]{gel1},
\begin{equation}\label{fourier}
\int_{-\infty}^{\infty} x^{-m} e^{i \sigma x}  \mathrm{d}x = i^m \frac{\pi}{(m-1)!} \sigma^{m-1} \mathrm{sgn}\sigma .
\end{equation}
With the identification $\sigma=(q-q')/\hbar$, we obtain the kernel of $\opr{p}^{-m}$,
\begin{eqnarray}\label{baba}
\kernel{\opr{p}^{-m}}=\frac{i (-1)^{\frac{1}{2}(m-1)}}{2\hbar^m (m-1)!} (q-q')^{m-1}\mbox{sgn}(q-q'), \;\;\; m=1, 2, \dots .
\end{eqnarray}
Substituting equation \eqref{baba} back into equation \eqref{wewe}, we obtain the kernel of the quantized operator $\opr{M}_{n,m}^{(Q)}$, 
\begin{equation}\label{quantize}
	\kernel{\opr{M}_{n,m}^{(Q)}} =  \frac{i (-1)^{\frac{1}{2}(m-1)}}{2\hbar^m (m-1)!} P_n(q|q')(q-q')^{m-1}\mbox{sgn}(q-q')
\end{equation}
where
\begin{equation}
	P_n(q|q') = \frac{1}{\sum_{j=0}^{n}a^{(n)}_{j}} \sum_{j=0}^{n} a^{(n)}_{j} q^{j} q'^{n-j} .
\end{equation}

The function $P_n(q|q')$ is a polynomial of order $n$ in both of its arguments. It satisfies the following properties: (i) $P_n(q|q')=P_n^*(q'|q)$, and (ii) $P_n(q|q)=q^n$. The first property enforces hermicity of the the quantized operator; and the second ensures that the classical observable emerges in the limit of vanishing $\hbar$. We will refer to such a polynomial as a quantizing polynomial. Any such polynomial serves as a quantization of the monomials $q^n p^{-m}$. Given a two-variable $n$-order polynomial, $p_n(q,q')$, satisfying all conditions for a quantizing polynomial, 
the corresponding quantized operator is given by 
\begin{eqnarray}
\opr{M}_{m,n}^{(Q_p)}=\sum_{j=0}^n \left[ \frac{1}{j!(n-j)!} \frac{\partial^{n} }{\partial q^j\partial q'^{n-j}} p_n(q,q')\right] \opr{q}^j \opr{p}^{-m} \opr{q}^{n-j} .
\end{eqnarray}

Now the associated quantizing polynomials for the Weyl, symmetric and Born-Jordan quantizations are given by
\begin{equation}\label{qfweyl}
P^{(W)}_n(q|q')=\frac{1}{\sum_{j=0}^n{n \choose j}}\sum_{j=0}^n{n \choose j} q^j q'^{n-j}=\frac{1}{2^n}\left(q+q'\right)^n ,
\end{equation}
\begin{equation}\label{qfsymmetric}
P^{(S)}_n(q|q')=\frac{1}{\sum_{j=0}^n (\delta_{0,j}+\delta_{n,j})}\sum_{j=0}^n (\delta_{0,j}+\delta_{n,j}) q^j q'^{n-j}=\frac{\left(q^n+q'^n\right)}{2} ,
\end{equation}
\begin{equation}\label{qfborn}
P_n^{(BJ)}(q|q')=\frac{1}{\sum_{j=0}^n 1}\sum_{j=0}^n q^j q'^{n-j}=\frac{1}{n+1}\frac{q^{n+1}-q'^{n+1}}{q-q'} ,
\end{equation}
obtained after performing the indicated summations. 

\section{Quantized Time of Arrival Operator}\label{qltoa}

\subsection{Quantizing the local time of arrival}

We now proceed in quantizing the local time of arrival for arbitrary potential and express the result in coordinate representation. Here we consider arrival at the origin. If the arrival point is different, then a simple translation can bring the problem to our treatment. We assume that the potential is analytic at the origin and thus admits the expansion $V(q)=\sum_{n=0}^{\infty} v_n q^n$. Under this condition, we can write
\begin{equation}\label{potdiff}
\int_0^q \left(V(q)-V(q')\right)^k \dd{q'} =  \sum_{n=1}^{\infty} a_n(k) q^n ,
\end{equation}
where the $a_n(k)$'s are constants independent of $q$. Then the local time of arrival assumes the form
\begin{equation}\label{ltoa2}
T=-\sum_{k=0}^{\infty}(-1)^k \frac{(2k-1)!!}{k!} \mu^{k+1} \sum_{n=1}^{\infty} a_n(k) q^n p^{-2k-1}.
\end{equation}
This form is now amenable to quantization. 

The local time of arrival is quantized by quantizing the monomial $q^n p^{-2k-1}$ for $n=1, 2, 3, \dots$ and $k=0, 1, 2, \dots$. Choosing a particular quantization of $q^n p^{-2k-1}$, $Q(q^np^{-2k-1})=\opr{M}_{n,2k+1}^{(Q)}$, the quantized LTOA is given by
\begin{equation}\label{ltoaq}
\opr{T}_Q=-\sum_{k=0}^{\infty}(-1)^k \frac{(2k-1)!!}{k!} \mu^{k+1} \sum_{n=1}^{\infty} a_n(k) \opr{M}_{n,2k+1}^{(Q)} ,
\end{equation}
where $Q$ indicates the chosen quantization. The kernel is 
\begin{eqnarray}
\kernel{\opr{T}_Q} &=&-\sum_{k=0}^{\infty}(-1)^k \frac{(2k-1)!!}{k!} \mu^{k+1} \sum_{n=1}^{\infty} a_n(k) \kernel{\opr{M}_{n,2k+1}^{(Q)}} 
\end{eqnarray}
From equation \eqref{generalquant}, the kernels of the quantized monomials are given by 
\begin{equation}
	\kernel{\opr{M}_{n,2k+1}^{(Q)}} =  \frac{i (-1)^k}{2\hbar^{2k+1} (2k)!} F^{(Q)}_n(q|q')(q-q')^{2k}\mbox{sgn}(q-q') .
\end{equation}
The kernel of the quantized-LTOA now assumes the form
\begin{eqnarray}
\kernel{\opr{T}_Q} = \frac{\mu}{i\hbar} \sum_{k=0}^{\infty} \frac{(2k-1)!!}{k! (2k)!} \frac{\mu^k (q-q')^{2k}}{2\hbar^{2k}}  J_k^{(Q)}(q,q') \sgn{q-q'}
\end{eqnarray}
where
\begin{equation}\label{jfq}
J_k^{(Q)}(q,q') = \sum_{n=1}^{\infty} a_n(k) F_n^{(Q)}(q|q') .
\end{equation}
We will see later that the $J_k(q,q')$'s assume a close form as integrals involving the potential, in particular, in terms of the integral \eqref{potdiff}.

The quantization of the local time of arrival is reduced to specifying the quantization function $F_n^{(Q)}(q|q')$ and evaluating the functions $J_k^{(Q)}(q|q')$. In accordance with our earlier results, we rewrite the kernel in the form
\begin{equation}\label{quantizedkernel}
\kernel{\opr{T}_Q} = \frac{\mu}{i\hbar} T_Q(q,q') \sgn{q-q'}
\end{equation}
\begin{equation}\label{kernelfq}
T_Q(q,q')=\frac{1}{2}\sum_{k=0}^{\infty} \frac{(2k-1)!!}{k! (2k)!} \frac{\mu^k (q-q')^{2k}}{\hbar^{2k}}  J_k^{(Q)}(q,q')
\end{equation}
All quantizations share this same functional form of the kernel. They only differ in the kernel factor $T_Q(q,q')$. Since $T_Q(q,q')$ is real valued, we have $\kernel{\opr{T}_Q}=\kernell{\opr{T}_Q}^*$ so that $\opr{T}_Q$ is hermitian; and also  $\kernel{\opr{T}_Q}^*=-\kernel{\opr{T}_Q}$ so that $\opr{T}_Q$ satisfies the time reversal symmetry $\opr{\Theta}\opr{T}_Q\opr{\Theta}^{\dagger}=-\opr{T}_Q$. By construction, $\opr{T}_Q$ takes the local time of arrival as its classical limit. Comparing these properties of the quantized LTOA-operators with the supraquantized TOA-operator, we see that the quantized operators differ only with the supraquantized operator in that they are not required to be conjugate with their respective Hamiltonians. However, as we have discussed in the Introduction, the Weyl quantized LTOA-operator coincide with the supraquantized TOA-operator for linear systems. In general, the quantized LTOA-operators will not be conjugate with their Hamiltonians, especially for non-linear systems.  

\subsection{Integral form of the kernel factors}
In this section we derive the explicit integral representations of the time kernel functions for the three most popular quantizations. The utility in the integral representation is not on that we can compute analytically the kernel factor from it. Its utility comes in handy when the kernel factor cannot be evaluated analytically and use the representation to compute the kernel by quadrature.

\subsubsection{Weyl quantization}
The kernel factor for Weyl quantization is obtained by first substituting the corresponding quantizing factor for $q^{n} p^{-m}$, $P^{(W)}_n(q|q')$, given by equation-\ref{qfweyl} back into equation-\ref{jfq} to obtain \begin{equation}\label{weyl}
J_k^{(W)}(q,q')=\sum_{n=1}^{\infty} a_n(k) \frac{(q+q')^n}{2^n}
\end{equation}
Comparing equation-(\ref{weyl}) with equation-\ref{potdiff}, we find that equation-\ref{weyl} is just equation-\ref{potdiff} with the replacement $q\rightarrow(q+q')/2$. Then we have
\begin{equation}\label{jfweyl}
J_k^{(W)}(q,q')=\int_0^{\frac{q+q'}{2}}\!\!\!\! \left(V\!\!\left(\frac{q+q'}{2}\right)-V(s)\right)^k\!\! \dd{s} .
\end{equation}
Then finally substituting $J_k^{(Q)}(q,q')$ back into equation-\ref{kernelfq} and interchanging the order of integration and summation, we obtain
\begin{equation}\label{kfweyl}
T_{W}(q,q')=\frac{1}{2} \int_0^{\frac{q+q'}{2}}\!\!\! ds\, _{0}F_{1}\! \left(;1; \frac{\mu}{2\hbar^2}(q-q')^2\left\{V\!\left(\frac{q+q'}{2}\right)-V(s)\right\}\right),
\end{equation}
in which $_{0}F_{1}$ is a particular hypergeometric function. In this form the earlier restriction on the potential can be lifted, provided of course that the integral exists. We have used earlier this representation in obtaining the quantized traversal time in the presence of potential barriers.

\subsubsection{Simple symmetric quantization}
Similarly employing the quantizing function for the simple symmetric ordering quantization given by equation-\ref{qfsymmetric}, we obtain 
\begin{eqnarray}\label{jfsymmetric}
J^{(S)}_k(q,q')&=&\frac{1}{2}\sum_{n=1}^{\infty} a_n(k) \left(q^n + q'^n\right)\nonumber\\
&=&\frac{1}{2}\left[\int_{0}^{q}(V(q)-V(s)^k \dd{s} + \int_{0}^{q'}(V(q')-V(s))^k\dd{s}\right]
\end{eqnarray}
Substituting this back into equation we obtain the time kernel factor for simple symmetric quantization and interchanging the order of integration and summation, we obtain
\begin{eqnarray}\label{kfsymmetric}
T_S(q,q') &=&  \frac{1}{4}\int_{0}^{q} \, _{0}F_{1}\! \left(;,1;\frac{\mu (q-q')^2 (V(q)-V(s))}{2\hbar}\right)\dd{s} \nonumber \\ 
&&+\frac{1}{4} \int_{0}^{q'}\, _{0}F_{1}\, \left(;,1;\frac{\mu (q-q')^2 (V(q')-V(s))}{2\hbar}\right) \dd{s}.
\end{eqnarray}

\subsubsection{Born-Jordan quantization}
Also using the quantizing function for the Born-Jordan quantization given by equation-\ref{qfborn}, we obtain
\begin{eqnarray}\label{jfborn}
J_k^{(BJ)}(q,q')&=&\frac{1}{(q-q')}\sum_{n=1}^{\infty} a_n(k)\frac{q^{n+1}-q'^{n+1}}{n+1}\nonumber \\
&=&\frac{1}{(q-q')}\sum_{n=1}^{\infty} a_n(k) \left[\int_0^q s^n \dd{s} - \int_0^{q'} s^n \dd{s}\right]\nonumber\\
&=&\int_{0}^{q} \int_{0}^{s} (V(s)-V(u))^k \dd{u} \dd{s} - \int_{0}^{q'}\int_{0}^{s}(V(s)-V(u))^k \dd{u} \dd{s} \nonumber
\end{eqnarray}
Similarly substituting this back yields the time kernel factor for the Born-Jordan quantization,
\begin{eqnarray}\label{kfborn}
T_{BJ}(q,q') &=& \frac{1}{2 (q-q')} \left[\int_{0}^{q}\int_{0}^{s}  \,_{0}F_1\!\left(;,1;\frac{\mu(q-q')^2}{2\hbar^2}(V(s)-V(u))\right) \dd{u}\dd{s}\right. \nonumber \\
&&\hspace{4mm}-\left. \int_{0}^{q'}\int_{0}^{s}  \,_{0}F_1\left(;,1;\frac{\mu(q-q')^2}{2\hbar^2}(V(s)-V(u))\right)\dd{u}\dd{s}\right] .
\end{eqnarray}

\section{The harmonic oscillator quantized local time of arrival operators}\label{harmonic}
We now apply our method to the harmonic oscillator. The potential energy function is given by $V(q)=\frac{1}{2}\mu \omega^2\,q^2$, where $\mu$ is the mass of the oscillator and $\omega$ is its angular frequency. Substituting the potential back into equation-\ref{prob}, we obtain the time of arrival at the origin,
\begin{equation}\label{harmonictoa}
T_0(q,p)=-\frac{1}{\omega}\tan^{-1}\left(\frac{\mu\omega\,q}{p}\right),
\end{equation}
where $q$ and $p$ are the initial position and momentum of the oscillator at time $t=0$, respectively. The classical time of arrival is multiple valued. This is due to periodicity of the motion of the classical harmonic oscillator, and it means that the particle will have multiple arrivals at the origin. However, as we have discussed above, this does not present any hindrance in the quantization of the LTOA of the harmonic oscillator. 

To obtain a quantization of the classical time of arrival for the first arrival at the origin, we expand equation-\ref{harmonictoa} about the free particle time of arrival to obtain the local time of arrival. The expansion yields the LTOA,
\begin{eqnarray}\label{harmonicltoa}
	\tau_{0}(q,p) = -\sum_{k=0}^{\infty} \frac{(-1)^{k}}{(2k+1)} \mu^{2k+1}\omega^{2k} \frac{q^{2k+1}}{p^{2k+1}} .
\end{eqnarray}
This expression is clearly real and single valued, and is convergent only in some small neighborhood of the origin for $p\neq 0$. It is this expression not equation-\ref{harmonictoa} that we subject to quantization. For a chosen quantization $Q$, the quantized local time of arrival assumes the form
\begin{eqnarray}
	\opr{T}_Q = -\sum_{k=0}^{\infty} \frac{(-1)^{k}}{(2k+1)} \mu^{2k+1}\omega^{2k} \opr{M}_{2k+1,2k+1}^{(Q)} .
\end{eqnarray}
The corresponding kernel of the quantized local time of arrival operator is now given by
\begin{eqnarray}
	\kernel{\opr{T}_Q} 
  = \frac{\mu}{i\hbar} T_Q(q,q') \sgn{q-q'}
\end{eqnarray}
where the kernel factor assumes the representation
\begin{equation}\label{harmonicfactor}
T_Q(q,q')=\frac{1}{2} \sum_{k=0}^{\infty}\frac{1}{(2k+1)!} \left(\frac{\mu\omega}{\hbar}\right)^{2k} P_{2k+1}^{(Q)}(q|q')(q-q')^{2k} .
\end{equation}
The explicit form of the kernel factor depends on the chosen quantization scheme. Also equation \eqref{harmonicfactor} can be summed explicitly without the need for the specific integral representation of the kernel factor.

Substituting the harmonic oscillator potential back into equations \eqref{kfweyl}, \eqref{kfsymmetric} and \eqref{kfborn}, we obtain the respective kernel factor for the Weyl, symmetric and Born-Jordan quantizations,
\begin{equation}\label{kfharmonicweyl}
T_W(q,q')=\frac{\hbar}{2\mu \omega (q-q')} \sinh\left(\frac{\omega \mu}{2\hbar}(q^2 - q'^2)\right).
\end{equation}
\begin{equation}\label{kfharmonicsymmetric}
T_S(q,q')=\frac{\hbar}{4\mu \omega (q-q')} \left[\sinh\left( \left( \frac{\mu \omega}{\hbar}\right)q(q-q') \right) + \sinh\left( \left( \frac{\mu \omega}{\hbar}\right)q'(q-q') \right) \right] .
\end{equation}
\begin{equation}\label{kfharmonicborn}
T_{BJ}(q,q')=\left(\frac{\hbar}{\mu \omega }\right)^2 \frac{1}{2(q-q')^3} \left[\cosh \left( \left( \frac{\mu \omega}{\hbar}\right)q(q-q') \right) - \cosh\left( \left( \frac{\mu \omega}{\hbar}\right)q'(q-q') \right) \right] ,
\end{equation}
We will be using these below to investigate the dynamics of the eigenfunctions of the quantized LTOA-operators.

It is straightforward to show that the kernel factor of the Weyl-quantized LTOA-operator, $T_W(q,q')$, solves the time kernel equation for the harmonic oscillator,
\begin{equation}
-\frac{\hbar^2}{2m}\frac{\partial^2 T(q,q')}{\partial q^2}+\frac{\hbar^2}{2 m} \frac{\partial^2 T(q,q')}{\partial q'^2} + \frac{m\omega^2}{2}(q^2-q'^2)T(q,q')=0
\end{equation}
and satisfies the required boundary conditions, $T(q,-q)=0$ and $T(q,q)=q/2$. This means that the Weyl-quantized LTOA-operator, $\opr{T}_W$, is conjugate to the harmonic oscillator Hamiltonian, which is a consequence of the linearity of the harmonic oscillator; and, in fact, coincides with the supraquantized TOA-operator operator for the harmonic oscillator \cite{galaponJMP2004}. 
On the other hand, the kernel factors $T_S(q,q')$ and $T_{BJ}(q,q')$ do not solve the time kernel equation so that the corresponding quantized LTOA-operators, $\opr{T}_S$ and $\opr{T}_{BJ}$, are not conjugate to the harmonic oscillator Hamiltonian.

\section{Generalized quantum images of the local time of arrival}\label{generalquantumimage}
Quantization does not exhaust all possible quantum images of the local time of arrival. However, the scalar form of quantization expressed in equation \eqref{quantize} points a way to extend quantization beyond the standard polynomial quantizations induced by ordering rules in constructing other quantum images of the monomials $q^n p^{-m}$. In its most general form, we can write the kernel of a quantum image of the monomial $q^n p^{-m}$ in the form
\begin{equation}\label{generalquant}
	\kernel{\opr{M}_{n,m}^{(F)}} =  F_n(q|q')\frac{i (-1)^{\frac{1}{2}(m-1)}}{2\hbar^m (m-1)!} (q-q')^{m-1}\mbox{sgn}(q-q'),
\end{equation}
where $F_n(q|q')$ is $C^{(\infty)}(\mathbb{R}\times\mathbb{R})$ in both its arguments and satisfies the conditions $F_n(q|q')=F_n^*(q'|q )$ and $F_n(q|q)=q^n$; moreover, $F_n(q|q')$ is independent of $\hbar$. The function $F_n(q|q')$ need not any more a polynomial in its arguments. We will refer to the function $F_n(q|q')$ as a quantum image (QI) function.

Now to show that equation-\ref{generalquant} is indeed a quantum image of $q^n p^{-m}$, we  show that $\opr{M}_{n,m}^{(F)}\rightarrow q^n p^{-m}$ in the classical limit $\hbar\rightarrow 0$. First, let us establish the expansion
\begin{equation}
F_n\!\left.\left(q+\frac{x}2\right|q-\frac{x}{2}\right) = q^n + \sum_{k=1}^{\infty} f_n(q)  x^k
\end{equation}
\begin{equation}
f_n(q)=\left.\frac{\partial^k F_n\!\left.\left(q+\frac{x}2\right|q-\frac{x}{2}\right)}{\partial x^k}\right|_{x=0}
\end{equation}
obtained by Taylor expansion about $x=0$. We substitute this back into the Weyl transform of $\kernel{\opr{M}_{n,m}^{(F)}}$ to obtain
\begin{equation}
\int_{-\infty}^{\infty}\transker{\opr{M}_{n,m}^{(F)}} e^{-i x p/\hbar} \, \mathrm{d}x =\frac{q^n}{p^m} + \sum_{k=1}^{\infty} \hbar^k\, \frac{f_n(q)}{p^{m+k}} .
\end{equation}
We arrived at this expression using the distributional Fourier transform 
\begin{equation}
\int_{-\infty}^{\infty} x^{m-1} \mathrm{sgn}(x) e^{-i x v} \mathrm{d}x = \frac{2(m-1)!}{i^{m} v^{m}}, 
\end{equation}
which is the inverse Fourier transform of equation-\ref{fourier}. 
Only the first term survives in the classical limit $\hbar\rightarrow 0$ so that $\opr{M}_{n,m}^{(F)}$ is indeed a quantum image of the classical monomial $q^n p^{-m}$. 

For a given quantum image of $q^n p^{-m}$ associated with some QI-function $F_n(q|q')$, we can obtain another quantum image by deforming $F_n(q|q')$. Let $\Omega(x)$ be a real valued $C^{(\infty)}(\mathbb{R})$ function with the properties $\Omega(0)=1$ and $\Omega(-x)=\Omega(x)$; $\Omega(x)$ may depend on $\hbar$ in the general case provided the dependence is such that the required classical limit is satisfied. Then we  obtain another quantum image with the QI-function
\begin{equation}
\tilde{F}_n(q|q')=F_n(q|q') \Omega(q-q').
\end{equation}
Consequently given a quantum image of the local time of arrival associated with the quantum image function $F_n(q|q')$, one obtains another quantum image with the corresponding kernel factor
\begin{equation}\label{newkernel}
\tilde{T}_{\tilde{F}}(q,q')=T_F(q,q')\Omega(q-q') .
\end{equation}
The corresponding operator to the kernel \eqref{newkernel} will have the local time of arrival as the classical limit. By a judicous choice of the function $\Omega(x)$, the time of arrival operator $\opr{T}_{\tilde{F}}$ is a bounded and self-adjoint operator in the Hilbert space $\mathcal{H}=L^2(\mathbb{R})$. For example, with an initial polynomial quantization, the choice $\Omega(x)=e^{-x^2/\sigma^2}$ leads to a bounded Fredholm type quantum image of the local time of arrival, say, for the harmonic oscillator. 

For the harmonic oscillator, both the Born-Jordan and simple-symmetric time kernel factors can be written as a deformation of Weyl time kernel factor. To do this, we use the addition and subtraction properties of hyperbolic function
\begin{equation}\label{sinh:identity}
\sinh x  + \sinh y= 2\sinh \left(\frac{x+y}{2}\right) \cosh\left(\frac{x-y}{2}\right) ,
\end{equation}
\begin{equation}\label{cosh:identity}
\cosh x - \cosh y = 2\sinh \left(\frac{x+y}{2}\right) \sinh \left(\frac{x-y}{2}\right).
\end{equation}
Applying equation \eqref{sinh:identity} on equation \eqref{kfharmonicsymmetric}, the kernel factor for the simple symmetric ordering, we obtain
\begin{eqnarray}
T_S(q,q)\!\!\!&=&\!\!\!\!\frac{\hbar}{2\mu \omega (q-q')} \left[\sinh\left( \left( \frac{\mu \omega}{2\hbar}\right)(q^2-q'^2) \right) \cosh\left( \left( \frac{\mu \omega}{2\hbar}\right)(q-q')^2 \right) \right]\nonumber \\
&=&T_W(q,q')\cosh\left( \left( \frac{\mu \omega}{2\hbar}\right)(q-q')^2 \right) 
\end{eqnarray}
Clearly $T_S(q,q')$ is a deformation of the Weyl time kernel  where the deformation function is given by 
$\Omega_S(x) = \cosh\left(\mu \omega x^2/2\hbar \right)$.
On the other hand, applying equation \eqref{cosh:identity} to simplify Born-Jordan kernel factor, we obtain
\begin{eqnarray}
T_{BJ}(q,q')&=&\left(\frac{\hbar}{\mu \omega }\right)^2 \frac{1}{(q-q')^3} \left[\sinh \left( \left( \frac{\mu \omega}{2\hbar}\right)(q^2-q'^2) \right) \sinh\left( \left( \frac{\mu \omega}{2\hbar}\right)(q-q')^2 \right) \right] \nonumber \\
&=& T_W(q,q') \frac{2\hbar}{\mu\omega(q-q')^2}\sinh\left( \left( \frac{\mu \omega}{2\hbar}\right)(q-q')^2 \right), 
\end{eqnarray}
This, too, is a deformation of the Weyl kernel factor with the deformation function $\Omega_{BJ}(x)=(2\hbar/\mu\omega x^2)\sinh\left(\mu \omega x^2/2\hbar \right)$. Notice that $\Omega(x)$ for these quantizations depend on $\hbar$.

\section{Time of arrival operators under parity transformation}\label{parity}
Now let us look into how a given TOA-operator $\opr{T}$ can posses eigenfunctions that have definite parities. Let $\opr{\Pi}$ be the parity operator defined by its action in position representation, $\opr{\Pi} \psi(q) = \psi(-q)$. The eigenfunctions of $\opr{T}$ have definite parities if the eigenfunctions are themselves eigenfunctions of $\Pi$. This follows provided $\opr{T}$ and $\opr{\Pi}$ commute.

We now obtain the sufficient condition for $\opr{T}$ and $\opr{\Pi}$ to commute. Applying the parity operator to both sides of $(\opr{T}\varphi)(q)=\int_{-\infty}^{\infty} \kernelpos{\opr{T}}\varphi(q')\,\mathrm{d}q'$, we obtain
\begin{equation}\label{eq:eigvalue2}
(\opr{\Pi}\opr{T}\varphi)(q)=\int_{-\infty}^{\infty}\kernelqq{-q}{\opr{T}}{q'}\varphi(q')\,\mathrm{d}q' 
\end{equation}
Performing the change in coordinate $q'\to -q'$ rewrites the foregoing equation into the form
\begin{eqnarray}\label{eq:eigvalue2}
(\opr{\Pi}\opr{T}\varphi)(q)=\int_{-\infty}^{\infty}\kernelneg{\opr{T}}\opr{\Pi}\varphi(q')\,\mathrm{d}q' ,
\end{eqnarray}
where we have implemented the definition of the parity operator. If it happens that the kernel is invariant under parity transformation in both of its arguments, i.e.
\begin{equation}\label{eq:Parity_condition}
\kernelneg{\opr{T}}=\kernel{\opr{T}},
\end{equation}
then we have the quality
\begin{equation}
(\opr{\Pi}\opr{T}\varphi)(q)=\int_{-\infty}^{\infty}\kernel{\opr{T}}\opr{\Pi}\varphi(q')\,\mathrm{d}q' = (\opr{T}\opr{H}\varphi)(q) .
\end{equation}
Since $\varphi(q)$ is arbitrary, this implies 
$\opr{\Pi} \opr{T} = \opr{T} \opr{\Pi}$ so that $\opr{T}$ and $\opr{H}$ commute when condition \eqref{eq:Parity_condition} is satisfied. Under such condition, the eigenfunctions of $\opr{T}$ have definite parities, i.e. either odd or even in $q$. 

We now look into the condition that leads to the invariance of the kernel under parity transformation. It is sufficient for us to consider the Weyl quantized LTOA-operator to demonstrate the analysis necessary to establish the condition. The kernel of the LTOA-operator is given by  
\begin{eqnarray}\label{eq:kernel_Weyl}
\kernel{\opr{T}_W}&=&\frac{\mu}{2i\hbar} \mbox{sgn}(q-q')\nonumber \\
&&\int_0^{\frac{q+q'}{2}}\!\!\! ds\, _{0}F_{1}\! \left(;1; \frac{\mu}{2\hbar^2}(q-q')^2 \left\{V\!\left(\frac{q+q'}{2}\right)-V(s)\right\}\right) .
\end{eqnarray}
Affecting the replacements $(q,q')\rightarrow (-q,-q')$ in the kernel \eqref{eq:kernel_Weyl}, followed by a change in variable $s\rightarrow -s$, the kernel transforms to
\begin{eqnarray}\label{eq:kernel_Weyl_neg}
\kernelneg{\opr{T}_W}
 &=&\frac{\mu}{2i\hbar} \mbox{sgn}(q-q') \nonumber \\
 &&\hspace{-14mm} \times \int_0^{\frac{q+q'}{2}}\!\!\! ds\, _{0}F_{1}\! \left(;1; \frac{\mu}{2\hbar^2}(q-q')^2 \left\{V\!\left(-\frac{q+q'}{2}\right)-V(-s)\right\}\right)\label{transWeyl}
\end{eqnarray}
Clearly when the potential is even, $V(q) = V(-q)$, equation \eqref{transWeyl} reduces to the untransformed kernel of the Weyl quantized TOA-operator \eqref{eq:kernel_Weyl}. Therefore, the Weyl-quantized TOA operator for even function potential commute with the parity, and thus possesses odd and even eigenfunctions.

Now when the potential is no longer symmetric, i.e. $V(-q)\neq V(q)$, the transformed kernel \eqref{transWeyl} does not reduce to the untransformed kernel \eqref{eq:kernel_Weyl}, so that the Weyl-quantized LTOA-operator does not commute with the parity operator. However, the RHS of equation \eqref{eq:kernel_Weyl_neg} is recognizably the kernel of a Weyl-quantized LTOA-operator corresponding to the interaction potential $\opr{\Pi}V(q)=V(-q)$. We define the TOA operators $\opr{T}_W^{\pm}$ where the positive sign is for $V=V(q)$ while  the negative sign is for $V=V(-q)$. We now wish to obtain the relationship between the eignfunctions of these two operators. Let $\varphi^+(q)$ be an eigenfunction of $\opr{T}_W^+$ with a corresponding eigenvalue $\tau$, i.e. $\opr{T}_W^+\varphi^+=\tau\varphi^+$. Then 
\begin{eqnarray}\label{eq:eigvalue}
\tau \opr{\Pi}\varphi^+(q)&=&\int_{-\infty}^{\infty}\kernelneg{\opr{T}^{+}_W}\opr{\Pi}\varphi^+(q')\,\mathrm{d}q' \nonumber \\
&=&\int_{-\infty}^{\infty}\kernel{\opr{T}^{-}_W}\opr{\Pi}\varphi^+(q')\,\mathrm{d}q' .
\end{eqnarray}
That is $\varphi^-=\Pi\varphi^+$ is an eigenfunction of $\opr{T}_W^-$ with the corresponding eigenvalue $\tau$. Conversely if $\varphi^-$ is an eigenfunction of $\opr{T}_W^-$ with an eigenvalue $\tau$, then $\Pi\varphi^-$ is an eigenfunction of $\opr{T}_W^+$ with the same eigenvalue $\tau$. 

The same analysis leads to the same conclusion for the simple symmetric and Born-Jordan quantizations of the local time of arrival.

\section{Dynamics of the Local Time of Arrival Operators}\label{dynamics}
We now look into the dynamics of the LTOA-operators. In particular, we wish to investigate the time development of their eigenfunctions in relation to their corresponding eigenvalues. In \cite{galaponPRA2005a,caballar2,caballarPRA2010} we found that when a free particle is prepared at time $t=0$ in a state which is an eigenfunction of its (free) time of arrival operator, the event that the position expectation value is equal to the arrival point and the event that the position uncertainty is minimum occur simultaneously at a later time equal to the eigenvalue corresponding to the eigenfunction. 
We referred to this as the unitary arrival of the eigenfunction at the arrival point, and it is this dynamical property that we expect and should demand from any legitimate time of arrival operator. Hence we devote the rest of the paper in investigating the dynamical behavior of the eigenfunctions of the Weyl, symmetric and Born-Jordan quantizations of the LTOA for the linear harmonic oscillator and for a non-linear potential. Also we will investigate the dynamics of the quantum images of the LTOA that are obtained by deforming a given quantization.

Our investigation will proceed in two steps: First is solving the eigenvalue problem for a given LTOA-operator; and second is evolving a given LTOA-operator eigenfunction by means of the time dependent Schrodinger equation. However, the eigenvalue problem is intractable analytically so we make headway by solving the problem numerically. The numerical solution is implemented by coarse graining the LTOA-operators by means of confining them in some closed interval in the real line, $[-l,l]$ for some $l>0$, and successively increasing $l$ to discern the behavior for arbitrarily large $l$. This leads to approximating the LTOA-operators by bounded and self-adjoint Fredholm integral operators, whose eigenvalue problems can be solved by quadrature. On the other hand, the Schrodinger equation is solved by replacing the kinetic energy with the kinetic energy of particle confined in the interval $[-l,l]$ under periodic boundary conditions, and by using the operator split method \cite{galaponIJMP2006}.

Our course graining is not a radical departure from the original configuration of the problem because, as we have discussed earlier, the LTOA-operator can be made bounded and self-adjoint by a judicious choice of the deformation function. The spatial confinement can interpreted as a deformation of the original operator in the same configuration space.

\subsection{The Quantized LTOA-operators for the Harmonic Oscillator}\label{sec:harmonic}
First, let us consider the quantized LTOA-operators of the harmonic oscillator corresponding to Weyl, Born-Jordan and simple symmetric orderings. The corresponding time kernel factors for each quantization are solved in Section-\eqref{harmonic}. We solve numerically the eigenfunctions of the QTOA operators in  Eq- \eqref{eq:eigvalue}, and evolve these with the parameters $\mu=\hbar=\omega=1$. We will compare the quantizations based on the expectation value and variance and the closeness of the time of occurrence of the quantum arrival to the corresponding eigenvalue. The eigenfunctions to be compared are those with eigenvalues that are approximately numerically equal.

\begin{figure}[t!]
    \centering
    \begin{subfigure}[b]{0.47\textwidth}
        \includegraphics[width=\textwidth]{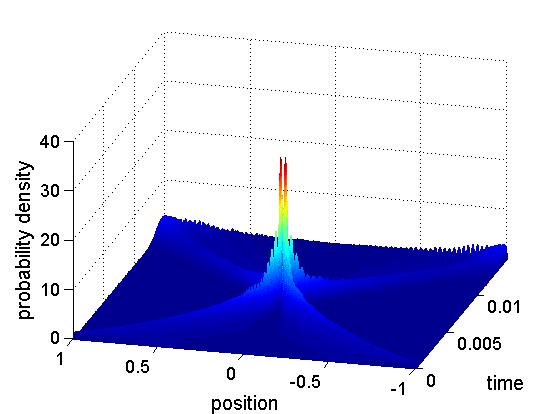}	   
        \label{fig:harmonic_quantize}
		\caption{}
\label{fig:harmonic_W_nodal}
    \end{subfigure}
    \begin{subfigure}[b]{0.47\textwidth}
        \includegraphics[width=\textwidth]{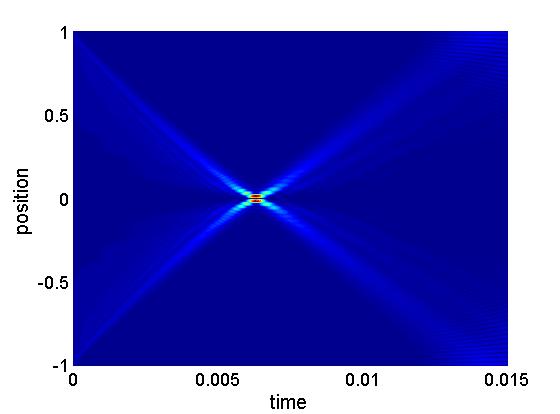}
        \caption{}
        \label{fig:harmonic_W_nodal_top}
    \end{subfigure}
    \begin{subfigure}[b]{0.47\textwidth}
        \includegraphics[width=\textwidth]{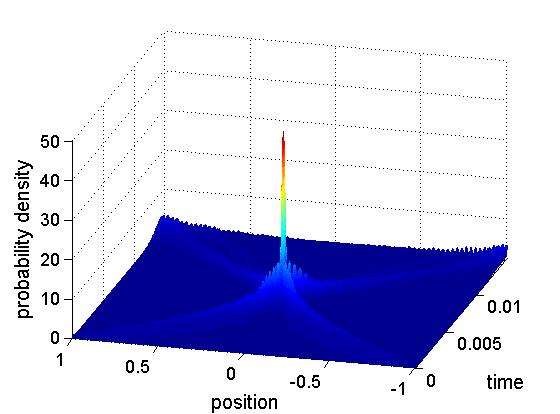}
        \caption{}
   \label{fig:harmonic_quantize}
        \label{fig:harmonic_W_antinodal}
    \end{subfigure}
    \begin{subfigure}[b]{0.47\textwidth}
        \includegraphics[width=\textwidth]{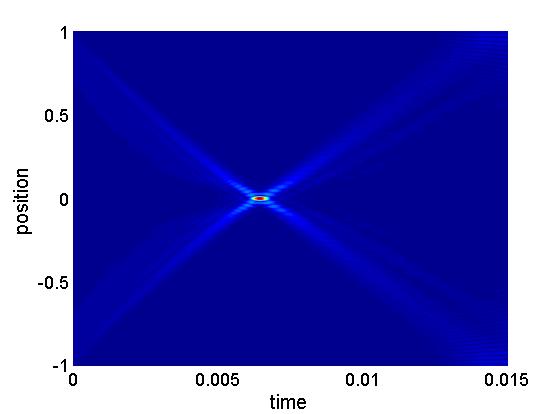}
        \caption{}
        \label{fig:harmonic_W_antinodal_top}
    \end{subfigure}
    \caption{Time evolution of the nodal (a,b) and (c,d) antinodal eigenfunctions of the Weyl-quantized TOA operator for harmonic oscillator. The eigenfunctions exhibit unitary collapse at their corresponding eigenvalue, which are 0.0063 and 0.00643 respectively.}
    \label{fig:harmonic_weyl}
\end{figure}

\begin{figure}[t!]
    \centering
    \begin{subfigure}[b]{0.47\textwidth}
        \includegraphics[width=\textwidth]{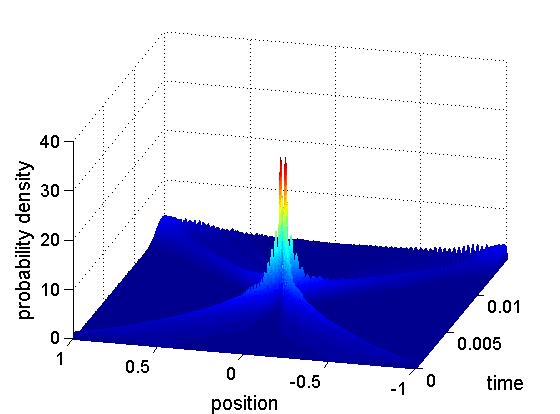}
 	   \caption{}\label{fig:harmonic_quantize}
        \label{fig:harmonic_S_nodal}
    \end{subfigure}
    \begin{subfigure}[b]{0.47\textwidth}
        \includegraphics[width=\textwidth]{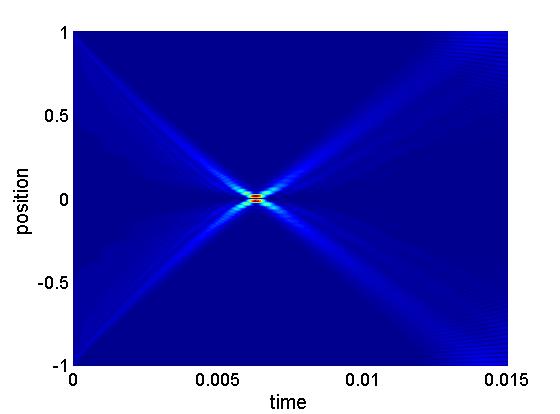}
        \caption{}
        \label{fig:harmonic_S_nodal_top}
    \end{subfigure}
    \begin{subfigure}[b]{0.47\textwidth}
        \includegraphics[width=\textwidth]{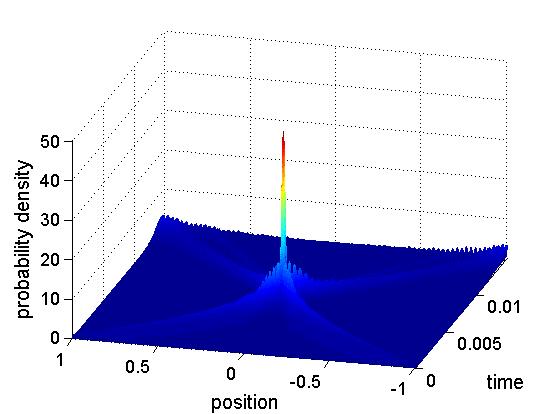}
        \caption{}
        \label{fig:harmonic_S_antinodal}
    \end{subfigure}
    \begin{subfigure}[b]{0.47\textwidth}
        \includegraphics[width=\textwidth]{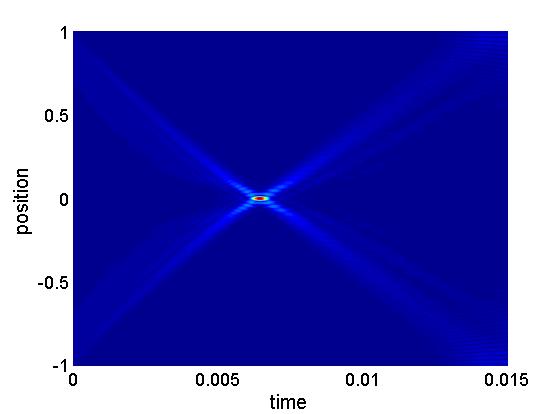}
        \caption{}
        \label{fig:harmonic_S_antinodal_top}
    \end{subfigure}
    \caption{Time evolution of the nodal (a,b) and (c,d) antinodal eigenfunctions of the Symmetric Ordering-quantized TOA operator for harmonic oscillator. The eigenfunctions exhibit unitary collapse at their corresponding eigenvalues which are 0.0063 and 0.00643 respectively.}\label{fig:harmonic_Symmetric}
\end{figure}

\begin{figure}[t!]
	\centering
   \begin{subfigure}[b]{0.47\textwidth}
    	\includegraphics[width=\textwidth]{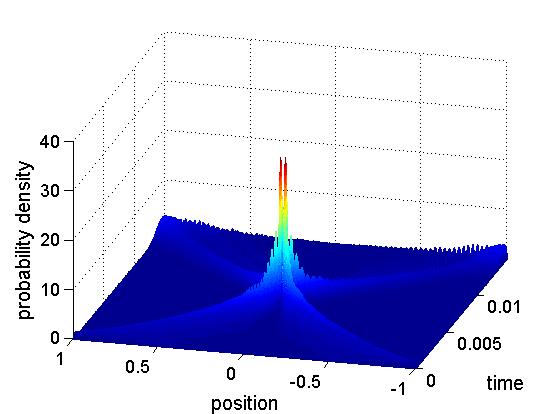}
   		\caption{}
        \label{fig:harmonic_B_nodal}
    \end{subfigure}    
    \begin{subfigure}[b]{0.47\textwidth}
        \includegraphics[width=\textwidth]{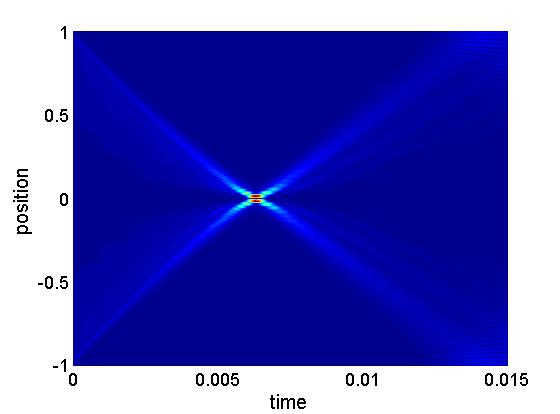}
        \caption{}
        \label{fig:harmonic_B_nodal_top}
    \end{subfigure}    
    \begin{subfigure}[b]{0.47\textwidth}
        \includegraphics[width=\textwidth]{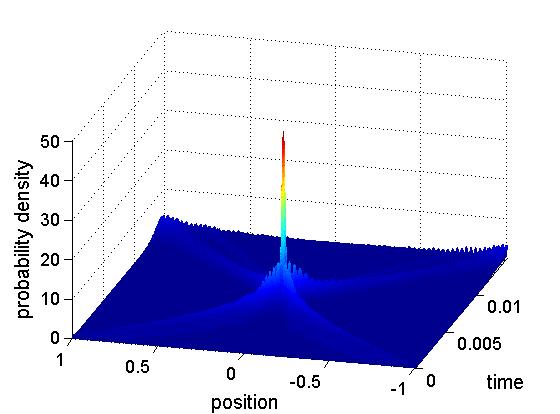}
        \caption{}
        \label{fig:harmonic_B_antinodal}
    \end{subfigure}    
    \begin{subfigure}[b]{0.47\textwidth}
        \includegraphics[width=\textwidth]{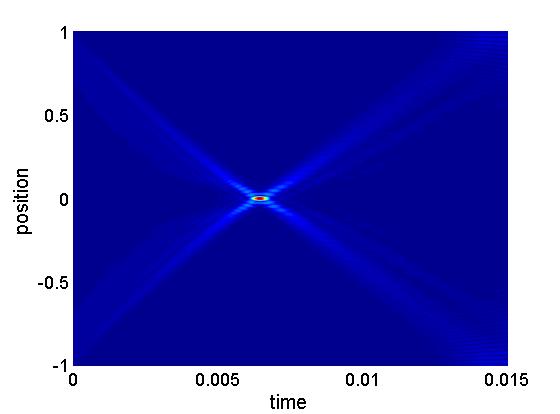}
        \caption{}
        \label{fig:harmonic_B_antinodal_top}
    \end{subfigure}    
    \caption{Time evolution of the nodal (a,b) and (c,d) antinodal eigenfunctions of the Symmetric Ordering-quantized TOA operator for harmonic oscillator. The eigenfunctions exhibit unitary collapse at their corresponding eigenvalues which are 0.0063 and 0.00643 respectively.}\label{fig:harmonic_Born_Jordan}
\end{figure}

Figures \ref{fig:harmonic_weyl}, \ref{fig:harmonic_Symmetric}, and \ref{fig:harmonic_Born_Jordan} show the representative dynamics of the eigenfunctions of the Weyl, simple symmetric and Born-Jordan quantized LTOA-operators for some fixed confining length $l$. The eigenfunctions (the wavefunctions at $t=0$ in the plots) are non-localized and are parity eigenfunction following from the even symmetry of the harmonic oscillator potential. Moreover, the eigenfunctions are either nodal or non-nodal in a similar fashion with the free time of arrival operator eigenfunctions. Nodal eigenfunctions have the characteristic dynamical property that two peaks coalesce at the arrival point (which is at the origin) with their closest approach to each other occurring at the respective eigenvalues of the eigenfunctions, after which the peaks disperse. On the other hand, the non-nodal eigenfunctions have the characteristic dynamical property that a single peak gathers at the origin with its minimum width occurring at the respective eigenvalues of the eigenfunctions, after which the peak disperses. Figure-4 shows the variance of the eigenfunctions as a function of time. In other words, the eigenfunctions of the quantized LTOA-operators unitarilly arrive at the origin at their respective eigenvalues within numerical accuracy. 

\begin{figure}[h!]
    \centering
    \begin{subfigure}[b]{0.47\textwidth}
        \includegraphics[width=\textwidth]{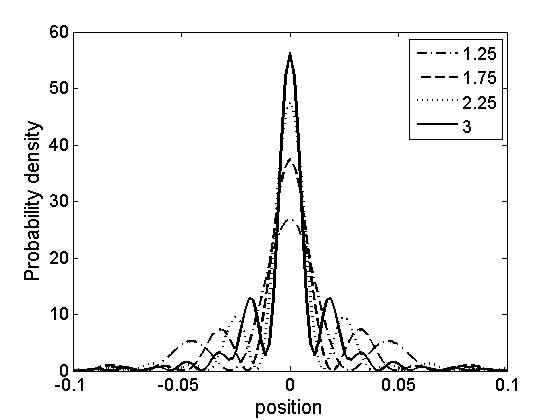}
        \caption{}
        \end{subfigure}
    \begin{subfigure}[b]{0.47\textwidth}
        \includegraphics[width=\textwidth]{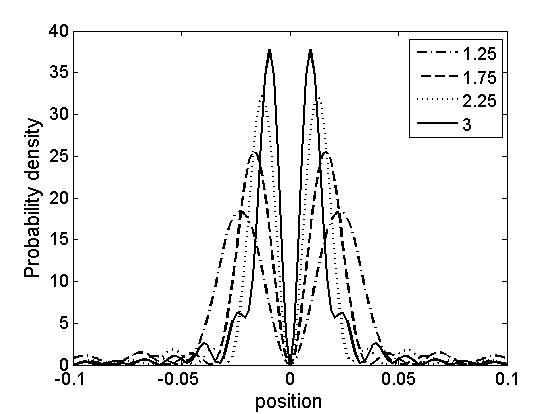}
        \caption{}
        \end{subfigure}
    \caption{Probability density $|\psi_n(q,\tau)|^2$ at time equal to the eigenvalue $\tau = 0.013$ for the antinodal (a) and nodal (b) eigenfunctions of Symmetric ordering quantized TOA operator for different values of $l$.}\label{fig:length_vary_Weyl}
\end{figure}

\begin{figure}[h!]
    \centering
    \begin{subfigure}[b]{0.47\textwidth}
    \includegraphics[width=\textwidth]{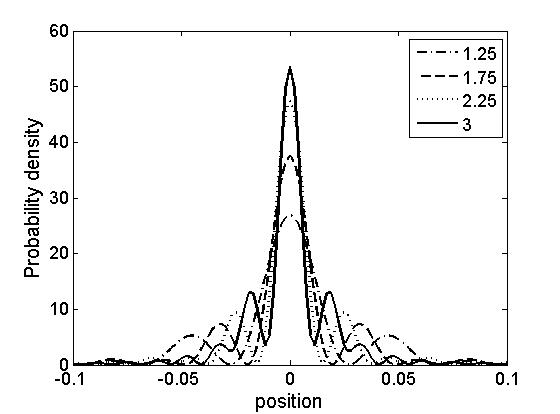}
        \caption{}
    \end{subfigure}
    \begin{subfigure}[b]{0.47\textwidth}
        \includegraphics[width=\textwidth]{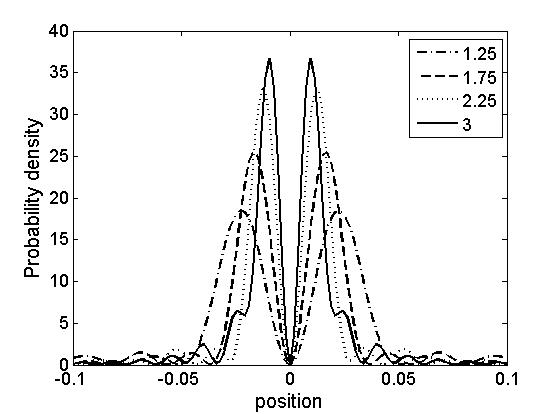}
        \caption{}
    \end{subfigure}
    \caption{Probability density $|\psi_n(q,\tau)|^2$ at time equal to the eigenvalue $\tau = 0.013$ for the antinodal (a) and nodal (b) eigenfunctions of Symmetric ordering quantized TOA operator for different values of $l$.}\label{fig:length_vary_Symmetric}
\end{figure}

\begin{figure}[h!]
    \centering
    \begin{subfigure}[b]{0.47\textwidth}
        \includegraphics[width=\textwidth]{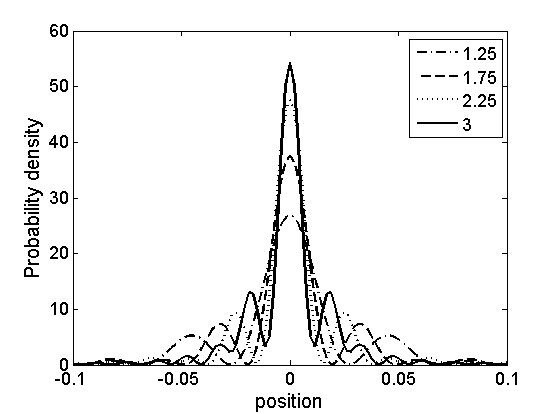}
        \caption{}
    \end{subfigure}
    \begin{subfigure}[b]{0.47\textwidth}
        \includegraphics[width=\textwidth]{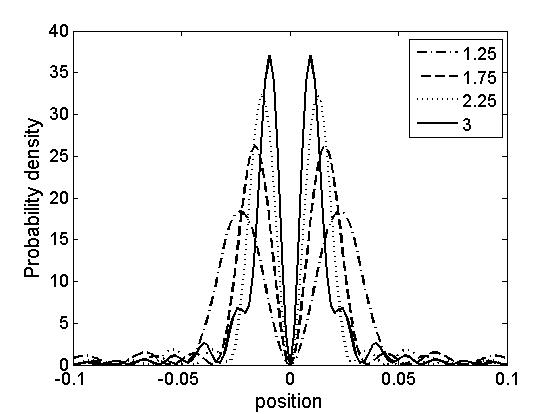}
        \caption{}
    \end{subfigure}
    \caption{Probability density $|\psi_n(q,\tau)|^2$ at time equal to the eigenvalue $\tau = 0.013$ for the antinodal (a) and nodal (b) eigenfunctions of Symmetric ordering quantized TOA operator for different values of $l$.}\label{fig:length_vary_Born_Jordan}
\end{figure}

Our foregoing results are for some fixed finite confining length $l$. To discern the dynamical behavior of the quantized LTOA-operators in the unbounded case, we successively increase $l$. Figures \ref{fig:length_vary_Weyl}, \ref{fig:length_vary_Symmetric} and \ref{fig:length_vary_Born_Jordan} show the plot of the probability density $|\psi _n (q,\tau _n)|^2$ of the evolved $n$-th eigenfunction $\psi_n(q)$ at time equal to the eigenvalue, $t=\tau_n$. For all quantizations considered, the probability density $|\psi _n (q,\tau _n)|^2$ becomes increasingly localized in the neighborhood of the arrival point as $l$ increases indefinitely.  Moreover, as the width  of $|\psi _n (q,\tau _n)|^2$ decreases with increasing $l$, the maximum value of $|\psi _n (q,\tau _n)|^2$ increases.  Inferring from the graphs, $|\psi _n (q,\tau _n)|^2$  will approach to a function with a singular support as $l$ approaches to infinity, consistent with our earlier observations \cite{galaponPRA2009,galaponJPA2009}. 

\begin{figure}[h!]
    \centering
    \begin{subfigure}[b]{0.47\textwidth}
        \includegraphics[width=\textwidth]{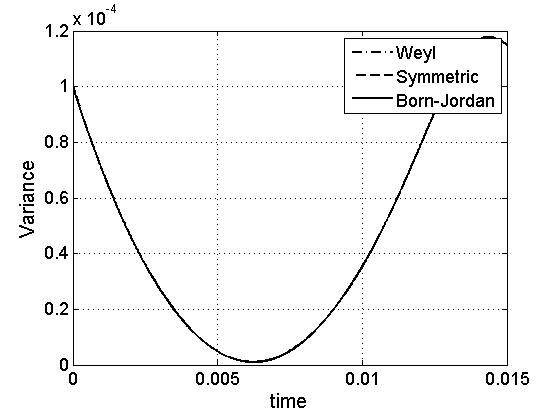}
        \caption{}
        \label{fig:harmonic_variance_nodal}
    \end{subfigure}
    \begin{subfigure}[b]{0.47\textwidth}
        \includegraphics[width=\textwidth]{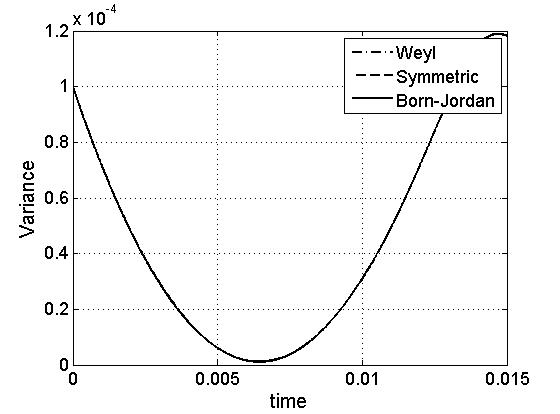}
        \caption{}
        \label{fig:harmonic_variance_antinodal}
    \end{subfigure}
    \caption{Position variance for nodal(a) and antinodal (b) eigenfunctions of QTOA operators for harmonic oscillator.}\label{fig:harmonic_variance}
\end{figure}

\subsection{Quantized time of arrival operators for a non-linear system}
We extend our analysis to non-linear systems, specifically to the system governed by the interaction potential 
\begin{equation}
V(q)=V_o \sin(a q),
\end{equation}
for some real constants $V_0$ and $a$.
We subject the local time of arrival for this potential to Weyl, symmetric and Born-Jordan quantizations. The corresponding kernel factors are obtained by means of the their integral representations given by Equations \eqref{kfweyl}, \eqref{kfsymmetric} and \eqref{kfborn} respectively. The kernel factors are now intractable to evaluate in closed form. So to proceed we evaluate them by quadrature. Since the potential is odd, the eigenfunctions are no longer eigenfunctions of the parity operator so that they do not have definite parities. 

\begin{figure}[h!]
    \centering
    \begin{subfigure}[b]{0.47\textwidth}
        \includegraphics[width=\textwidth]{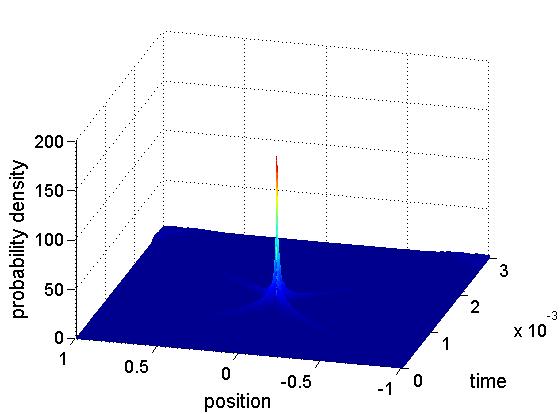}
    \caption{}\label{fig:nodal_gen_quantize}
    \label{fig:gen_W_nodal}
    \end{subfigure}
    \begin{subfigure}[b]{0.47\textwidth}
        \includegraphics[width=\textwidth]{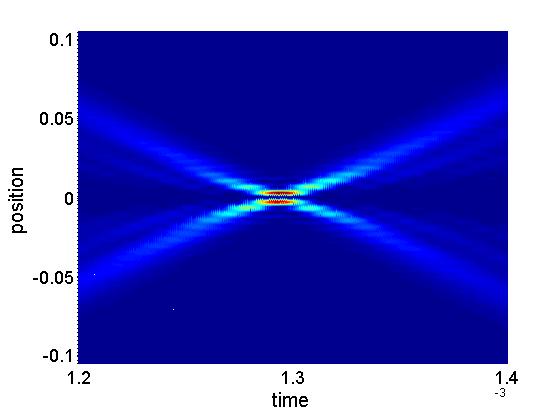}
        \caption{}
        \label{fig:gen_W_nodal_top}
    \end{subfigure}
    \begin{subfigure}[b]{0.47\textwidth}
        \includegraphics[width=\textwidth]{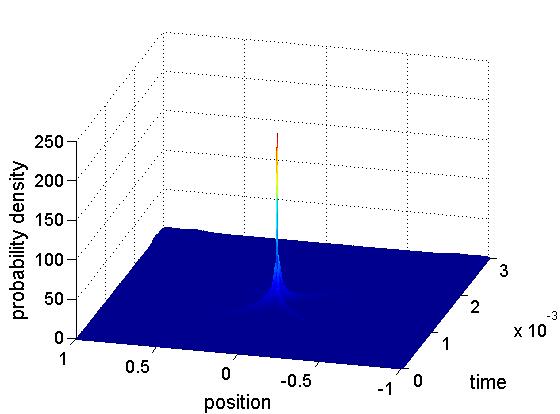}
        \caption{}
        \label{fig:gen_W_antinodal}
    \end{subfigure}
    \begin{subfigure}[b]{0.47\textwidth}
        \includegraphics[width=\textwidth]{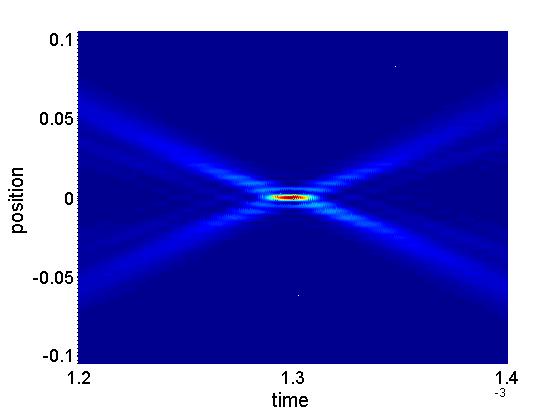}
        \caption{}
        \label{fig:gen_W_antinodal_top}
    \end{subfigure}
    \caption{Time evolution of the nodal (a,b) and antinodal (c,d) eigenfunctions of the Weyl quantized TOA operator for $V(q) = \sin (q)$. The eigenfunctions exhibit unitary collapse at the arrival point at their corresponding eigenvalues which are $1.287\times 10^{-3}$ and $1.292 \times 10^{-3}$ respectively.}  \label{fig:gen_Weyl}
\end{figure}

\begin{figure}[h!]
    \centering
    \begin{subfigure}[b]{0.47\textwidth}
        \includegraphics[width=\textwidth]{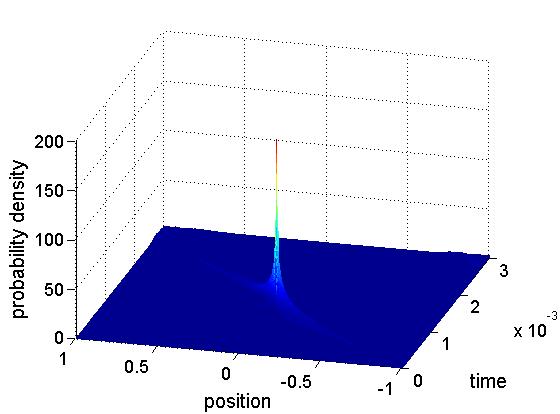} 
        \caption{}
        \label{fig:gen_S_nodal}
        \end{subfigure}
    \begin{subfigure}[b]{0.47\textwidth}
        \includegraphics[width=\textwidth]{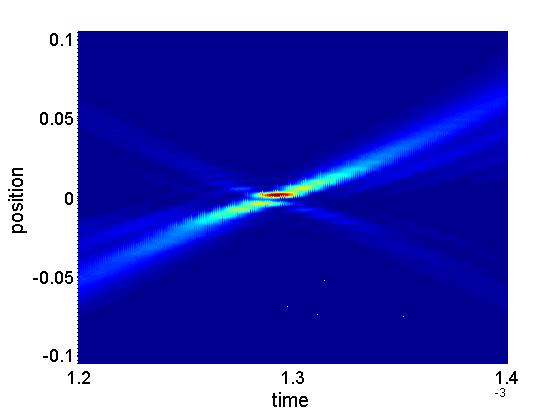}
        \caption{}
        \label{fig:gen_S_nodal_top}
    \end{subfigure}
       \begin{subfigure}[b]{0.47\textwidth}
        \includegraphics[width=\textwidth]{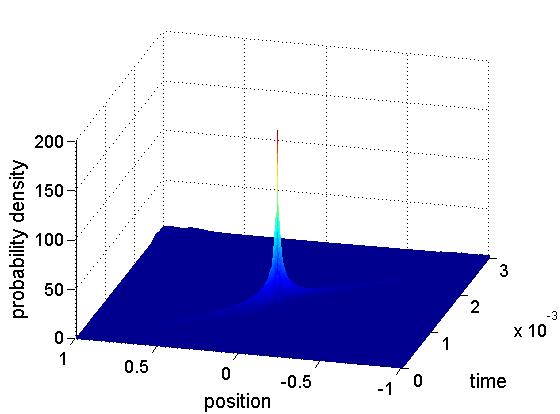}
        \caption{}
        \label{fig:gen_S_antinodal}
    \end{subfigure}
    \begin{subfigure}[b]{0.47\textwidth}
        \includegraphics[width=\textwidth]{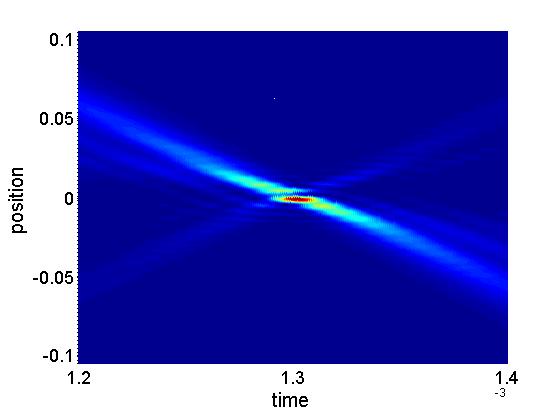}
        \caption{}
        \label{fig:gen_S_antinodal_top}
    \end{subfigure}
    \caption{Time evolution of the nodal (a,b) and antinodal (c,d) eigenfunctions of the Symmetric Ordering-quantized TOA operator for $V(q) = \sin (q)$. The eigenfunctions exhibit unitary collapse at the arrival point at their corresponding eigenvalues, which are $1.285\times 10^{-3}$ and $1.295\times 10^{-3}$ respectively.}\label{fig:gen_Symmetric}
\end{figure}

\begin{figure}[h!]
    \centering
    \begin{subfigure}[b]{0.47\textwidth}
        \includegraphics[width=\textwidth]{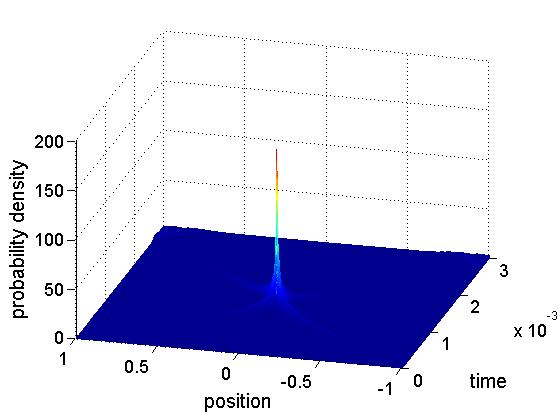}
    \caption{}
        \label{fig:gen_B_nodal}
    \end{subfigure}
    \begin{subfigure}[b]{0.47\textwidth}
        \includegraphics[width=\textwidth]{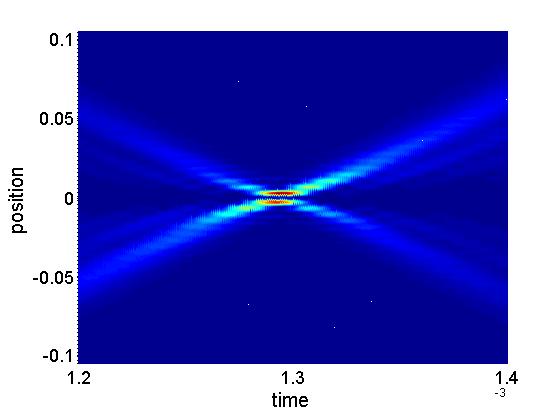}
        \caption{}
        \label{fig:gen_B_nodal_top}
    \end{subfigure}
    \begin{subfigure}[b]{0.47\textwidth}
        \includegraphics[width=\textwidth]{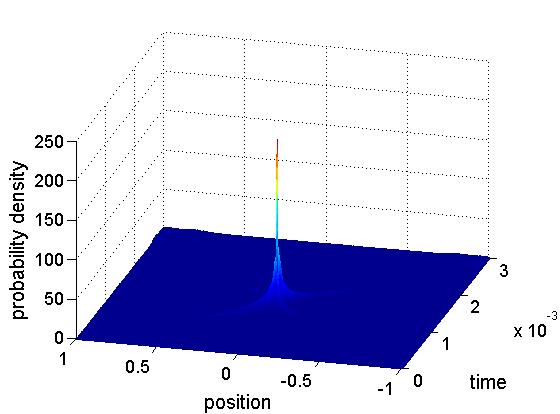}
        \caption{}
        \label{fig:gen_B_antinodal}
    \end{subfigure}
    \begin{subfigure}[b]{0.47\textwidth}
        \includegraphics[width=\textwidth]{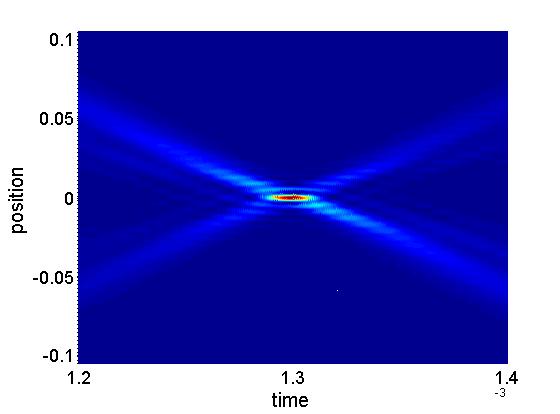}
        \caption{}
        \label{fig:gen_B_antinodal_top}
    \end{subfigure}
    \caption{Time evolution of the nodal (a,b) and antinodal (c,d) eigenfunctions for Born - Jordan quantized TOA operator of $V(q) = \sin (q)$. The eigenfunctions exhibit unitary collapse at the arrival point at their corresponding eigenvalues, which are $1.287 \times 10^{-3}$ and $1.292 \times 10^{-3}$ respectively.}\label{fig:gen_Born_Jordan}
\end{figure}

\begin{figure}[h!]
    \centering
    \begin{subfigure}[b]{0.47\textwidth}
        \includegraphics[width=\textwidth]{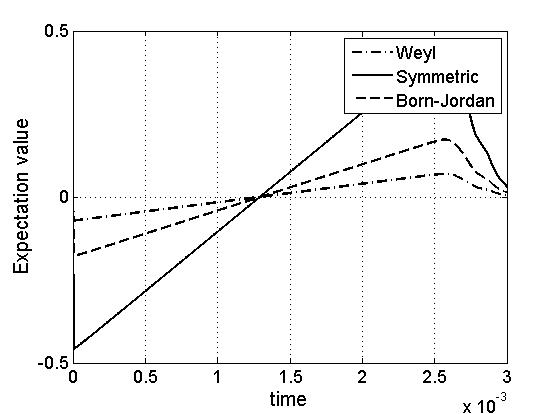}
        \caption{}
        \label{fig:gen_exp_nodal_orig}
    \end{subfigure}
    \begin{subfigure}[b]{0.47\textwidth}
        \includegraphics[width=\textwidth]{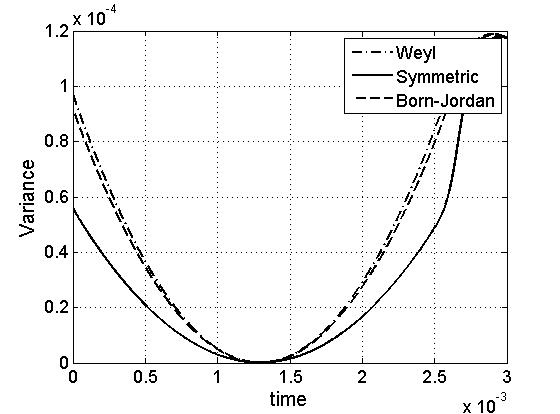}
        \caption{}
        \label{fig:gen_expval_nodal_closer}
    \end{subfigure}
    \caption{Expectation value (a) and variance (b) of the position through time for the nodal eigenfunction of TOA operator for $V(q) = \sin (q)$ }\label{fig:gen_graph_nodal}
\end{figure}

\begin{figure}[h!]
    \centering
    \begin{subfigure}[b]{0.47\textwidth}
        \includegraphics[width=\textwidth]{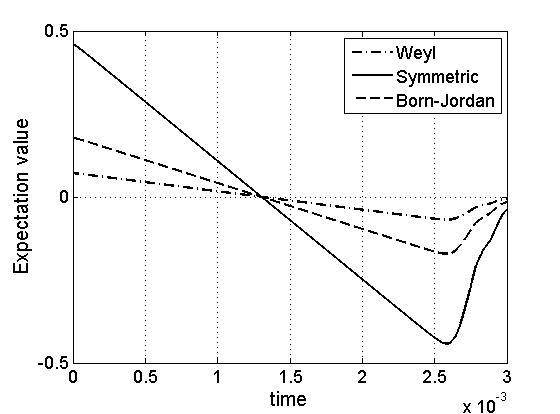}
        \caption{}
    \end{subfigure}
    \begin{subfigure}[b]{0.47\textwidth}
        \includegraphics[width=\textwidth]{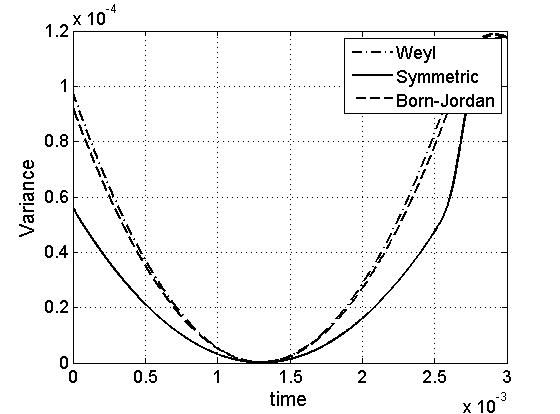}
        \caption{}
    \end{subfigure}
    \caption{Position expectation value (a) and variance (b) for different time values of the antinodal eigenfunction of the TOA operator for $V(q) = \sin (q)$. The times at which the expectation value is equal to the arrival point and  at which the position variance is minimum, are equal to the corresponding  }\label{fig:gen_graph_antinodal}
\end{figure}

Figures \ref{fig:gen_Weyl}, \ref{fig:gen_Symmetric}, \ref{fig:gen_Born_Jordan},  \ref{fig:gen_graph_nodal}  and  \ref{fig:gen_graph_antinodal} show the representative dynamics of the eigenfunctions of the LTOA-operators for the three quantizations. Again we observe that the eigenfunctions unitarilly arrive at the origin at their respective eigenvalues. We have, in particular, the position expectation value with respect to the eigenfunctions arriving at the intended arrival point at the corresponding eigenvalues. 

However, we can differentiate the quantizations by comparing the symmetry of probability density about the arrival point. By comparing the magnitude of the slope of the position expectation value in Figures \ref{fig:gen_graph_nodal} and \ref{fig:gen_graph_antinodal}, we see that the most asymmetric probability density is that of the simple symmetric quantized TOA while the most symmetric is that of the Weyl quantized TOA operator. In fact, as the asymmetry increases, the eigenfunction types becomes less of a nodal-antinodal eigenfunction. If we prefer the quantization that have nodal and antinodal eigenfunctions, we would prefer the Weyl quantization over the two other quantization. 


\subsection{Deformed quantized local time of arrival operators}
\begin{figure}[h!]
    \centering
    \begin{subfigure}[b]{0.47\textwidth}
        \includegraphics[width=\textwidth]{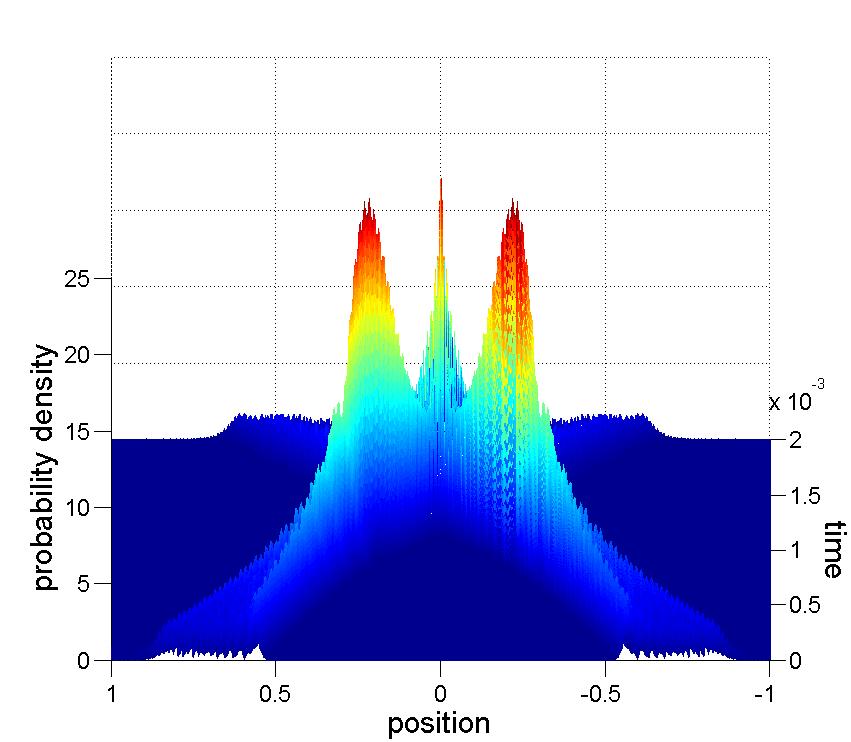}
        \caption{}
    \end{subfigure}
    \begin{subfigure}[b]{0.47\textwidth}
        \includegraphics[width=\textwidth]{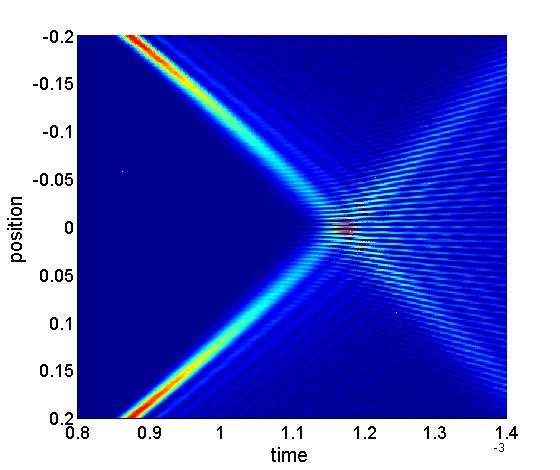}
        \caption{}
    \end{subfigure}
    \caption{Time evolution of an eigenfunction of a TOA that was obtained by multiplying the Weyl time kernel by $\Omega(q-q') = 1 + 20000(q-q')^2$. Its corresponding eigenvalue is $1.06 \times 10^{-3}$.}\label{fig:trans_quant}
\end{figure}

The quantized TOA-operators so far exhibit unitary arrival of their eigenfunctions at their respective eigenvalues. We now consider deformations of the quantized LTOA-operators. Again these operators have the same properties and classical limits, which is the classical local time of arrival. We consider the LTOA-operator arising from the deformed Weyl LTOA-operator for harmonic oscillator with deformation function $\Omega(q-q') = 1 + \alpha (q-q')^2$. FOr small values of $\alpha$, the eigenfunctions exhibit unitary arrival at the intended arrival point at their respective eigenvalues within numerical accuracy; however, as the value of $\alpha$ increases, the unitary arrival becomes less sharp until unitary arrival is no longer discernible, as shown in Figure \ref{fig:trans_quant} for a large value of $\alpha$. More important, the probability density does not take its maximum value at the  the eigenvalue and does not occur in the neighborhood of the arrival point. 

\section{Conclusion}\label{conclusion}
In this paper we considered the quantization of the classical time of arrival observable. We have argued that there is nothing in principle in the laws of quantum mechanics that forbids the quantization of the classical time of arrival. The earlier objections to the quantization due to possible multivaluedness and complexity of the classical time arrival have been dealt with with the quantization of the local time of arrival. The different (Weyl, simple-symmetric, and Born-Jordan) quantized LTOA-operators are compared by means of the dynamical behaviors of their eigenfunctions. We observed, within the parameters and accuracy of our numerical simulations, that the eigenfunctions evolve such that the event that the position expectation value is equal to the arrival point and the event that the position uncertainty is minimum occur simultaneously at times equal to the corresponding eigenvalues of the eigenfunctions. That is the eigenfunctions unitarilly arrive at the predetermined arrival point at their respective eigenvalues. This leads to the conclusion that the Weyl, symmetric, and Born-Jordan quantized LTOA-operators are legitimate time of arrival operators.  

However, our observations raise the fundamental question of the role played by the conjugacy or non-conjugacy of the quantized LTOA-operators with the Hamiltonian in their dynamics. It is reasonable to suspect that somehow conjugacy has a role in the dynamics of the quantized LTOA-operators, especially in the observed unitary arrival of their eigenfunctions at the arrival point. But our numerical simulations show that Weyl, symmetric and Born-Jordan quantizations manifest the same unitary arrival at the arrival point. However, they do not simultaneously satisfy the conjugacy relation with the Hamiltonians. For example, for linear systems, only the Weyl-quantized LTOA-operators are conjugate with the Hamiltonian. A reasonable hypothesis is that the operator that satisfies the conjugacy relation will have the sharpest arrival at the arrival point. By ``sharpest arrival'' we mean that (i) the event of the arrival of the position expectation value at the arrival point and the event that the position uncertainty is minimum occur {\it exactly} at the same time equal to the eigenvalue, and (ii) the manifested minimum uncertainty is already the smallest possible value.  We have, in particular, the SQTOA-operator in mind as the TOA-operator exhibiting sharpest arrival. A counter  hypothesis is that it is the classical limit and not conjugacy with the Hamiltonian that determines the dynamics so that it is expected that all quantizations behave similarly because they have the same classical limit. 

The observed similarities in the dynamics of the Weyl, Symmetric and Born-Jordan quantizations of the local time of arrival seems to support the counter hypothesis. However, our last result on the deformed Weyl quantized LTOA-operator for the harmonic oscillator does not support the counter hypothesis. There we observed that the eigenfunctions exhibit unitary arrival, within numerical accuracy, at the arrival point for sufficiently small values of the parameter $\alpha$. However, the observed unitary arrival steadily degrades as the value $\alpha$ increases until no unitary arrival is no longer observed. Specifically, for sufficiently large $\alpha$, the evolving eigenfunctions still manifest tendency to localize but the peaks of the probability density do not occur at the corresponding eigenvalues nor at the arrival point. Since the deformed LTOA-operator has the local time of arrival as its classical limit, this result already roles out the classical limit as the sole determinant of the dynamical behavior of the eigenfunctions of the quantized LTOA-operators. But how about the observed similarities in the behavior of the Weyl, Symmetric and Born-Jordan quantizations? We have shown above that the Symmetric and the Born-Jordan quantizations are deformations or perturbations of the Weyl quantization of the local time of arrival. Their observed similarities may indicate they introduce small perturbations to the Weyl-quantized LTOA-operator that no significant deviation from the dynamical behavior of Weyl quantized operator can be observed. This expectation is not unreasonable because, as we have just noted, the behavior of the deformed Weyl-quantized operator for sufficiently small $\alpha$ behaves similarly with that of the Weyl-quantized operator. This may be the case for the Symmetric and Born-Jordan quantized operators for linear systems.  

In general, the SQTOA-operator has an expansion the leading term of which is the Weyl quantization of the local time of arrival, with the succeeding terms introducing  quantum corrections of order $O(\hbar^2)$ in the classical limit. In this regard Weyl quantization takes on a special role as it determines the leading term of the time of arrival operator conjugate to the Hamiltonian. In fact we can, too, cast the other quantizations such that the leading term is just the Weyl quantization of the classical observable, as exemplified by the Symmetric and Born-Jordan quantizations. This allows us to look at the other quantizations as perturbations of Weyl quantization. Given that the Weyl quantized LTOA-operator is the leading term in any quantized LTOA-operator, should we expect that the Weyl quantized LTOA-operator manifest the sharpest unitary arrival in non-linear systems among all possible quantizations of the local time of arrival? If conjugacy determines the dynamics of the time of arrival operators, then it is not unreasonable to anticipate the existence of a quantization or quantizations that will lead to sharper arrivals than Weyl quantization. That is because quantizations other than Weyl will have additional terms that may get the quantized LTOA -operator closer to the SQTOA-operator. However, it is possible, too, that the additional terms may possibly perturb the Weyl-quantized LTOA-operator farther away from the SQTOA-operator that the Weyl-quantized operator remains to exhibit sharpest arrival among possible quantizations for non-linear systems. Clearly much remains to be done to understand the exact role of the canonical commutation relation in the dynamics of the quantum time of arrival operators.  

\section*{Acknowledgement} This work was supported by the Depart Science of Science and Technology through the National Research Council of the Philippines.

\end{document}